\documentclass{aa}  

\usepackage{graphicx}
\usepackage{txfonts}
\usepackage{color}
\usepackage{bm}
\begin{document} 

   \title{SIP-IFVM: Efficient time-accurate magnetohydrodynamic model of the corona and coronal mass ejections}
   \author{H. P. Wang
          \inst{1,2}
          \and
          J. H. Guo\inst{1,3}
          \and
          L. P. Yang\inst{2}
          \and
          S. Poedts\inst{1,4}
          \and
          F. Zhang\inst{5,6}
          \and
          A. Lani\inst{1,7}
          \and 
          T. Baratashvili\inst{1}
          \and
          L. Linan\inst{1}
          \and
          R. Lin\inst{1,8}
          \and Y. Guo\inst{3}
          }
   \institute{Centre for Mathematical Plasma-Astrophysics, Department of Mathematics, KU Leuven, Celestijnenlaan 200B,
      3001 Leuven, Belgium\\
\email{haopeng.wang1@kuleuven.be}
\and
State Key Laboratory of Space Weather, Chinese Academy of Sciences, Beijing 100190, China
\and
School of Astronomy and Space Science and Key Laboratory of Modern Astronomy and Astrophysics, Nanjing University, Nanjing 210023, China\\     
\email{jinhan.guo@kuleuven.be}
\and 
Institute of Physics, University of Maria Curie-Skłodowska, ul. Radziszewskiego 10, 20-031 Lublin, Poland\\
\email{Stefaan.Poedts@kuleuven.be}
\and
Institute of Theoretical Astrophysics, University of Oslo, PO Box 1029 Blindern, 0315 Oslo, Norway   
\and
Rosseland Centre for Solar Physics, University of Oslo, PO Box 1029 Blindern, 0315 Oslo, Norway       
\and
Von Karman Institute For Fluid Dynamics, Waterloosesteenweg 72, 1640 Sint-Genesius-Rode, Brussels, Belgium
\and
School of Earth and Space Sciences, Peking University, Beijing 100871, China}


 
  \abstract
   { Coronal mass ejections (CMEs) are one of the main drivers of space weather. However, robust and efficient numerical modelling applications of the initial stages of CME propagation and evolution process in the sub-Alfv{\'e}nic corona are still lacking.}
   {Magnetohydrodynamic (MHD) solar coronal models are critical in the Sun-to-Earth model chain, but they do sometimes encounter low-$\beta$ ($<10^{-4}$)  problems near the solar surface. This paper aims to deal with these low-$\beta$ problems and make MHD modelling suitable for practical space weather forecasting by developing an efficient and time-accurate MHD model of the solar corona and CMEs. In this paper, we present an efficient and time-accurate three-dimensional (3D) single-fluid MHD solar coronal model and employ it to simulate CME evolution and propagation.}
   {Based on a quasi-steady-state implicit MHD coronal model, we  developed an efficient time-accurate coronal model that can be used to speed up the CME simulation by selecting a large time-step size. We have called it the Solar Interplanetary Phenomena-Implicit Finite Volume Method (SIP-IFVM) coronal model. A pseudo-time marching method was implemented to improve temporal accuracy. A regularised Biot-Savart Laws (RBSL) flux rope, whose axis can be designed into an arbitrary shape, was inserted into the background corona to trigger the CME event. We performed a CME simulation on the background corona of Carrington rotation (CR) 2219 and evaluated the impact of time-step sizes on simulation results. Our study demonstrates that this model is able to simulate the CME evolution and propagation process from the solar surface to $20\; R_s$ in less than 0.5 hours (192 CPU cores, $\sim$ 1 M cells). Compared to the explicit counterpart, this implicit coronal model is not only faster, but it also has improved numerical stability. We also conducted an ad hoc simulation with initial magnetic fields artificially increased. It shows that this model can effectively deal with time-dependent low-$\beta$ problems ($\beta<10^{-4}$). Additionally, an Orszag-Tang MHD vortex flow simulation demonstrates that the pseudo-time-marching method used in this coronal model can simulate small-scale unsteady-state flows. }
   {The simulation results show that this MHD coronal model is very efficient and numerically stable. It is a promising approach to  simulating time-varying events in the solar corona with low plasma $\beta$ in a timely and accurate manner.}
   {}


   \maketitle
%

\section{Introduction}
   Disastrous space weather is caused by solar storms that propagate to the Earth's orbit and impact space- and ground-based infrastructures that are vital to modern society. Thus, there is an urgent need to understand the mechanism of space weather and to eventually  gain access to reliable forecasts hours to days in advance \cite[e.g.][]{BAKER19987,Feng_2011Chinese,Feng_2013Chinese,Feng2020book,Koskinen2017,Poedts_2020}. To achieve this goal, we need to develop and improve advanced numerical models to better understand the complex mechanisms of space weather.
   
   Physics-based MHD modelling is a first principal method capable of bridging the large heliocentric distance from near the Sun to well beyond Earth's orbit self-consistently and revealing the fundamental propagation and evolution processes of solar storms \cite[e.g.][]{Detman2005,Dryer2007nodoc,Feng_2007,Feng_2010,Feng_2011,Feng_2012,Feng2012,FENG20141965,Feng_2014,Feng_2017,Feng2020book,Gombosi2018,Hayashi2006,Lihuichao2018,LiHuichao2020,LUGAZ20111187,Mikic1999,Nakamizoetal2009,RILEY20121,Shen_2021,TOTH2012870,Usmanov1993,Usmanov2003,WuShiTsan2015,Yang_2021,Zhou2012,Zhouyufen2017}. However, realistic MHD simulation is a complex process involving disparate physical spatiotemporal scales and is computationally intensive. To predict severe space weather events in timely
and accurately way,  efficient, spatiotemporal accurate, and numerical stable MHD models are anticipated \cite[e.g.][and references therein]{Feng2020book,Owens2017}.  
   
   Usually, coupling different models dedicated to specific regions and physical problems is the preferred choice for establishing a space-weather forecasting framework \cite[e.g.][]{Feng_2013Chinese,GOODRICH20041469,Hayashi_2021,Kuzma_2023,ODSTRCIL20041311,Perri_2022,Perri_2023,Poedts_2020,TOTH2012870}. Among these coupled components, the solar coronal model is essential for initialising the other models and plays a key role in influencing the simulation results of solar disturbance propagation and evolution \citep{Brchnelova_2022,Perri_2023}. In the solar corona, the solar wind velocity increases from subsonic to supersonic. Also, Solar disturbances, such as coronal mass ejections (CMEs) and solar proton events, propagate through the solar corona \citep{Feng2020book,Kuzma_2023}.     
  
   However, MHD simulations of the solar corona are also the most complex and computationally intensive component and sometimes encounter low $\beta$ (the ratio of the thermal pressure to the magnetic pressure) problems with $\beta$ as low as $10^{-4}$ near the solar surface \citep{Bourdin2017}. For example, to update the solution for 1 hour of physical time requires about 50~hrs of computing time on $100\;$CPUs in the data-driven MHD modelling of a flux-emerging active region inside the low corona \citep{Jiang201605}. In this simulation, $\beta=2\times10^{-3}$, with a time step size smaller than $0.1$ seconds due to the restriction of Courant-Friedrichs-Lewy (CFL) stability condition. Even a steady-state global solar coronal simulation by an explicit MHD model, using a solenoidality-preserving approach to maintain magnetic field divergence-free constraints, takes about $50\;$hrs of computing time on $576$ MPI processes to obtain a steady-state solution \citep{FengandLiu2019}. In this simulation, $\beta=1\times 10^{-3}$ exists near the solar surface, and the time-step size is consequently limited to around $3.6$ seconds.  
   
   Since a high level of efficiency and numerical stability are required in practical applications, compromises are usually made in solar coronal simulations. For instance, empirical Wang-Sheeley-Arge \citep[WSA;][]{Arge2003ImprovedMF,Yangzicai2018} solar coronal model in EUHFORIA \citep{Poedts_2020, Pomoell2018020}, the magneto-frictional (MF) coronal nonlinear force-free fields module in MPI-AMRVAC \citep{Guo_2016}, and the simplified physics-based zero-$\beta$ solar coronal model \citep{Caplan_2019} are used to estimate the solar coronal structures. However, these simplifications discard a lot of important information and it was demonstrated that an MHD model provides better forecasts than an empirical solar coronal model \citep{Samara_2021}. Therefore, many researchers are devoted to establishing more efficient and accurate MHD solar coronal models \citep{Feng2020book, Kuzma_2023}.

   Recently, \cite{Feng_2021} and \cite{Wang_2022,Wang2022_CJG} have established a series of second-order accurate, efficient, and robust implicit MHD solar coronal models capable of dealing with low-$\beta$ problems. They reduced the wall-clock times of steady-state MHD solar coronal simulations from several days to less than 1 hour under the same computing environment. However, these steady-state models still have significant potential for improvement in spatiotemporal resolution. In this paper, we further  develop a time-accurate and numerical stable implicit MHD solar coronal model, calling it  Solar-Interplanetary Phenomena-Implicit Finite Volume Method (SIP-IFVM) coronal model. The SIP-IFVM coronal model is employed to simulate the evolution and propagation procedure of CMEs in the background corona. Based on the work presented in this paper, we will further develop spatiotemporal high-order accurate and computationally efficient implicit MHD solar coronal models in the future.

   In the implicit algorithm, although the convergence rate can be improved by selecting a considerable time step \citep{brchnelova2023role,Feng_2021,Kuzma_2023,Liu_2023,Perri_2022,Perri_2023,WANG201967,Wang_2022,Wang2022_CJG}, it may be accompanied by a loss in temporal accuracy \citep{Linan_2023}. By modelling the evolution and propagation of flux ropes, it was proven that the implicit solar coronal model can be time-accurate and still faster than the explicit MHD model by selecting a suitable time step size \citep{guo2023,Linan_2023}. Besides, some researchers employed the pseudo-time marching method by introducing a pseudo time $\tau$ at each physical time step and solving a steady-state problem on $\tau$ to guarantee the temporal accuracy \cite[e.g.][]{BIJL2002313,Lingquan2019,LUO2001137,SITARAMAN2013364}. However, rough selections of the physical time step or the pseudo-time step may reduce computation efficiency. To make the time-accurate simulation more efficient, some flexible time-step size adaption strategies \citep{Hoshyari_2020,NOVENTA2020104529} are proposed and implemented in fluid dynamics simulations of both steady and unsteady flows. In this paper, we first employ the implicit MHD model with a considerable time step to model the steady-state coronal structures.  In the next step, we  select a relatively small time step and carry out the pseudo-time marching method at each physical time step to guarantee both computational efficiency and the necessary accuracy for CME simulations.

   After establishing this efficient time-accurate coronal model capable of dealing with low $\beta$ problems, we carried out CME simulations to further validate the model's capability of modelling the propagation and evolution processes of CMEs. Generally, two kinds of models exist to initiate a CME in a numerical space-weather framework. The first one is based on the non-magnetic hydrodynamic cloud, such as the plasma-sphere cone model \citep{Hayashi2006CONEFR,Mays2015,Odstrcil1999,Pomoell2018020,Zhao2002}. This model can roughly retrieve the geometry and dynamics of CMEs, namely, the angular width, ejection direction and propagation speed. Therefore, this model is generally used to predict the arrival time of CMEs and the intensity of their induced shock. However, observations suggest that most CMEs include a magnetic flux rope. For instance, filaments \citep{GUO202108,Ouyang2017} and hot channels \citep{Chen2017,Zhang2012} are often found to be progenitors of CMEs, which are frequently served as the proxies of magnetic flux ropes in observations. Many non-linear-force-free-field extrapolation and data-driven models also demonstrate the existence of pre-eruptive flux ropes \citep{Cheung2012,Guo2017,Pomoell2018020}. Furthermore, observations of white-light coronagraph observations found that almost one-half of the CMEs manifest a twisted flux rope structure \citep{Vourlidas2013}. In addition, some in situ detections in interplanetary space found that CMEs usually hold a monotonic rotation of internal magnetic fields, indicating the twisted flux-rope field lines \citep{Burlaga1981}. As a result, constrained by these observations, it seems that magnetic flux rope-based CME-initialisation models are more realistic. 

   In recent decades, many magnetic flux rope models used to initiate a CME have been proposed. Among these, torus-shaped and cylindrical flux rope models are widely used to initialise CME simulations \citep{Kataoka2009,Marubashi2016,Nieves_Chinchilla_2018,Singh_2020,Scolini2019,Yang_2021,Zhang2019,Zhouyufen2017}. Such as the Gibson-Low flux rope model \citep{Gibson_1998} adopted in the AWSoM model \citep{Jin_2017}. Also, some analytically modified Titov-D\'{e}moulin circular \citep{Titov_2014} and 'S-shaped' regularised Biot-Savart laws \citep{Titov_2018} flux ropes have been implemented in COCONUT (COolfluid COroNal UnsTructured) coronal model \citep{guo2023,Linan_2023}, MAS code \citep{Linker2024}, PLUTO code \citep{Regnault_2023}, and MPI-AMRVAC \citep{Guo2019,Keppens2023}. In this paper, we further adopt the RBSL flux rope, which allows  the electric current path to take on an arbitrary
shape, with an aim to initialise CME simulations in a numerically stable, efficient, and time-accurate implicit thermodynamic MHD coronal model. What makes this SIP-IFVM model different from the aforementioned coronal models is mainly its adoption of the parallel LU-SGS implicit solver, the pseudo-time marching method, the approximate linearisation strategy, the decomposed MHD equation, and the six-component grid system described below. These combined features significantly enhance the coronal model's efficiency, time accuracy, and numerical stability.

   Firstly, we model a CME evolution and propagation process driven by an RBFL flux rope with a theoretical `S'-shape curve path to validate the novel algorithms proposed in this paper. Secondly, we demonstrate the model's capability of modelling a robust magnetic environment by an ad hoc simulation with enhanced background corona and flux rope magnetic fields. We also perform an Orszag-Tang MHD vortex flow simulation \citep{orszag_tang_1979} to show that the novel pseudo-time-marching method adopted in this paper is capable of simulating small-scale unsteady-state flows, so does this MHD coronal model. 
   Considering that there is still a lack of robust and efficient modelling of the initial stages of CME propagation and shock evolution in the sub-Alfv{\'e}nic corona below about 20 $R_s$ \citep{Vourlidas2019}, the present coronal model is what is needed to play an active role in improving space weather forecasting capabilities.

   This paper is organised as follows. In Sect.~\ref{sec:Description of the novel MHD Model}, the numerical formulation and implementations of the MHD coronal model are described in detail, including the discretisation of the governing equations,  implementation of the pseudo-time marching method, and  processing of boundary conditions. We demonstrate the simulation results in Sect.~\ref{sec:Numerical Results} and Appendix A. In Sect.~\ref{sec:conclusion}, we summarise the main features of the efficient time-accurate implicit MHD coronal model and give the concluding remarks.

\section{Governing equations and numerical methods}\label{sec:Description of the novel MHD Model}
\subsection{Governing equations}\label{The governing equations} 
We solved the decomposed MHD equations where the magnetic field $\mathbf{B}=\left(B_x,B_y,B_z\right)^T$ is split into a time-independent potential magnetic field, $\mathbf{B}_0=\left(B_{0x},B_{0y},B_{0z}\right)^T$, and a time-dependent field $\mathbf{B}_1=\left(B_{1x},B_{1y},B_{1z}\right)^T$ \cite[e.g.][]{Feng_2010,FUCHS2010JCP,Guo2015,Powell1999,Tanaka1995,Wang_2022}. The governing equations are the same as those in \cite{Wang_2022} and are described in the following compact form:
\begin{equation}\label{MHDinsolarwind}
\frac{\partial \mathbf{U}}{\partial t}+\nabla \cdot \mathbf{F}\left(\mathbf{U}\right)=\frac{\partial \mathbf{U}}{\partial t}+\frac{\partial \mathbf{f}\left(\mathbf{U}\right)}{\partial x}+\frac{\partial \mathbf{g}\left(\mathbf{U}\right)}{\partial y}+\frac{\partial \mathbf{h}\left(\mathbf{U}\right)}{\partial z} =\mathbf{S}\left(\mathbf{U},\nabla \mathbf{U}\right).\
\end{equation}
   Here, $\mathbf{U}=\left(\rho, \rho \mathbf{v}, E_1, \mathbf{B}_1\right)^T$ denotes the conservative variable vector, $\nabla \mathbf{U}$ corresponds to the derivative of $\mathbf{U}$, $\mathbf{F}\left(\mathbf{U}\right)=\left(\mathbf{f},\mathbf{g},\mathbf{h}\right)$ is the inviscid flux vector with $\mathbf{f}$, $\mathbf{g}$, and $\mathbf{h}$ denoting the components in the $x$, $y$, and $z$ directions, and $\mathbf{S}=\mathbf{S}_{\rm Powell}+\mathbf{S}_{\rm gra}+\mathbf{S}_{\rm rot}+\mathbf{S}_{\rm heat}$ represents the source term vector including the Godunov-Powell source terms, the gravitational force, the Coriolis force, and the heating source terms. 
   
   The governing equations described above are used to conduct simulations for the steady-state background solar corona and the evolution and propagation of CMEs. First, we performed a time-relaxation procedure to achieve a quasi-steady background corona. When the steady-state simulation converges, we added the flux rope magnetic field $\mathbf{B}_{FR}(\mathbf{x})$ calculated by the RBSL flux rope model \cite{guo2023} to $\mathbf{B}_{1}$ of the quasi-steady background corona.We then carried out the subsequent time-accurate CME simulation.

  \subsection{Solver description}\label{sec:Description of the RBF-based FR method}
   In this paper, we adopted Godunov's method to advance cell-averaged solutions in time by solving a Riemann problem at each cell interface \citep{EINFELDT1991273,Godunov1959Adifference}. By integrating Eq.~(\ref{MHDinsolarwind}) over the hexahedral  cell $i$ and using Gauss's theorem to calculate the volume integral of the divergence of flux, we reach the following discretised integral equations:
\begin{equation}\label{MHDequationdiscrization}
V_{i}\frac{\partial \mathbf{U}_{i}}{\partial t}=-\oint_{\partial V_{i}}\mathbf{F}\cdot \mathbf{n} d \Gamma+V_{i}\mathbf{S}_{i},\
\end{equation}
   where $\oint_{\partial V_{i}}\mathbf{F}\cdot \mathbf{n} d \Gamma=\sum\limits_{j=1}^{6}\mathbf{F}_{ij}\cdot \mathbf{n}_{ij} \Gamma_{ij}$ and $\mathbf{S}_i=\mathbf{S}_{{\rm Powell},i}+\mathbf{S}_{{\rm gra},i}+\mathbf{S}_{{\rm rot},i}+\mathbf{S}_{{\rm heat},i}$. Then, $\mathbf{U}_{i}$ and $\mathbf{S}_{i}$ refer to the cell-averaged solution variable and source term in cell $i$, $V_{i}$ is the volume of cell $i$, $\Gamma_{ij}$ means the area of interface $i,j$ shared by cell $i$ and its neighbouring cell $j$, and $\mathbf{n}_{ij}$ is the unit normal vector of $\Gamma_{ij}$ and points from cell $i$ to cell $j$. As in \cite{Feng_2021} and \cite{Wang_2022}, $\mathbf{S}_{{\rm gra},i}$ and $\mathbf{S}_{{\rm rot}, i}$ are calculated as the corresponding variables at the centroid of cell $i$, $\mathbf{S}_{{\rm Powell},i}$ is calculated by employing Gauss's law and mean value theorem.
   The inviscid flux through the interface $\Gamma_{ij}$, $\mathbf{F}_{ij}\cdot \mathbf{n}_{ij}=\mathbf{F}_{ij}\left(\mathbf{U}_{L},\mathbf{U}_{R}\right) \cdot \mathbf{n}_{ij}$, is calculated by the positive-preserving HLL Riemann solver \citep{Feng_2021}. The subscripts `$_{L}$' and `$_{R}$' denote the corresponding variables on $\Gamma_{ij}$ extrapolated from cell $i$ and cell $j$, respectively. 
    Additionally, the cell-averaged heat source terms, $\mathbf{S}_{{\rm heat},i}=\left(0,\mathbf{S}_{m,i},Q_{e,i}+\mathbf{v}_i\cdot\mathbf{S}_{m,i}+\left(\nabla\cdot\mathbf{q}\right)_i,\mathbf{0}\right)^T$, are calculated in a similar way as did in \cite{Wang_2022}. It means that $\mathbf{S}_{m,i}$, $Q_{e,i}$, and $\mathbf{v}_i$ are defined as the corresponding variables at the centroid of cell $i$ and $\left(\nabla\cdot\mathbf{q}\right)_i$ is calculated by Green-Gauss method, $\left(\nabla\cdot\mathbf{q}\right)_i=\frac{1}{V_{i}} \sum\limits_{j=1}^{6}  \mathbf{q}_{ij} \cdot \mathbf{n}_{ij} \Gamma_{ij}$, where $\mathbf{q}_{ij}\left(T_{L},T_{R},\left.\left(\nabla T\right)\right|_i,\left.\left(\nabla T\right)\right|_j,\mathbf{U}_{L},\mathbf{U}_{R}\right)$ is the heat flux through $\Gamma_{ij}$. 

   The state variables, as well as the derivatives of temperature on the cell surface, $\Gamma_{ij}$, are required in calculating the inviscid flux and heat conduction term through $\Gamma_{ij}$.
   For convenience, we utilised a second-order positivity-preserving reconstruction method to calculate the piecewise polynomials of primitive variables as:
\begin{equation}\label{FlowfieldbyLSQ}
X_i(\mathbf{x})=\left.X\right|_i+\Psi_i\left.\nabla X\right|_i\cdot\left(\mathbf{x}-\mathbf{x}_i\right)
,\end{equation}
   where $X \in \{\rho,u,v,w,p\}$, $\left.X\right|_i$ is the corresponding variable at $\mathbf{x}_i$, the centroid of cell $i$, and $\left.\nabla X\right|_i=\left.\left(\frac{\partial X}{\partial x},\frac{\partial X}{\partial y},\frac{\partial X}{\partial z}\right)\right|_i$ is the derivative of $X$ at $\mathbf{x}_i$. $\Psi_i$ is the limiter used to control spatial oscillation. Meanwhile, the temperature at the cell centroid is derived from equation of state $\left.T\right|_i=\frac{\left.p\right|_i}{\Re \left.\rho\right|_i}$ and the reconstruction formulation of temperature in cell $i$, denoted by $T_i(\mathbf{x})$, is also calculated by Eq.~(\ref{FlowfieldbyLSQ}). A discontinuity detector \citep{Feng_2021,Li2012ijnm} is used to determine whether the Barth-limiter \citep{Barth1989} should be triggered to control spatial oscillation for $\rho,u,v,w, p,T$.

   For the magnetic field, a globally solenoidality-preserving (GSP) approach \citep{FengandLiu2019,Feng_2021} is employed to maintain the divergence-free constraint, $\frac{1}{V_i} \sum\limits_{j=1}^{6}\mathbf{B}_{ij} \cdot \mathbf{n}_{ij} \Gamma_{ij}=0$, by performing an iteration procedure when reconstructing $\mathbf{B}_0$ and $\mathbf{B}_1$ on the cell interface. However, applying a limiter to the magnetic field can compromise its divergence-free constraint and may also increase discretisation error in the magnetic field. 
   Considering that $\left(\mathbf{B}+\epsilon~\mathbf{B}\right)^2-\mathbf{B}^2\equiv 2~\epsilon~\mathbf{B}^2+\epsilon^2~\mathbf{B}^2$, the discretisation error in magnetic pressure caused by the increased discretisation error of magnetic field, denoted by $\epsilon~\mathbf{B}$, can be comparable to thermal pressure and non-physical negative thermal pressure may appear when deriving thermal pressure from energy density, especially in low $\beta$ regions.  
   Therefore, in addition to utilising the decomposed MHD equations, we discarded the limiter for the magnetic field during the quasi-steady coronal simulation to avoid any degradation in the accuracy and divergence-free constraint on the magnetic field. However, in the following time-varying CME simulations, a limiter for the magnetic field is still necessary to control spatial oscillation. In this work, we adopted a continuously differentiable WBAP limiter \citep{Feng2020book,Li2011,Li2012} for $\mathbf{B}_1$ (as described in Eq.~\ref{WBAP}) and constrained the max iteration for GSP to 5 as a prelimary attempt. We found it worked well in our CME simulations.
These equations are expressed as:
\begin{equation}\label{WBAP}
\begin{aligned}
X_i(\mathbf{x})&=\left.X\right|_i+ 
\left.\begin{pmatrix}
\psi_{i,1}^{\rm WBAP}\frac{\partial X}{\partial x}\\
\psi_{i,2}^{\rm WBAP}\frac{\partial X}{\partial y}\\
\psi_{i,3}^{\rm WBAP}\frac{\partial X}{\partial z}
\end{pmatrix}\right|_i\cdot\left(\mathbf{x}-\mathbf{x}_i\right),\\
\text{ with }\psi_{i,\xi}^{\rm WBAP}&\left(1,\vartheta_1,\vartheta_2,\cdots,\vartheta_6\right)=\\
&\left\{\begin{array}{c}
\frac{n+\sum\limits_{j=1}^6\frac{1}{\vartheta_{j,\xi}^{q-1}+\epsilon}}{n+\sum\limits_{j=1}^6\frac{1}{\vartheta_{j,\xi}^{q}+\epsilon}}, \text { if } \vartheta_1,\vartheta_2,\cdots,\vartheta_6 > 0\\
0,  \text { else }
\end{array}\right.,\
\end{aligned}
\end{equation}
   where $\vartheta_{j,\xi}=\left.\frac{\partial X}{\partial\xi}\right|_j\big/\left.\frac{\partial X}{\partial\xi}\right|_i$ with $j\in\{1,2,\cdots,6\}$, $\xi \in\{x,y,z\}$ and $X \in\{B_{1x},B_{1y},B_{1z}\}$. $\epsilon$, $n$, and $q$ are adjustable parameters and we set $\epsilon=10^{-8}$, $n=1$, and $q=4$ in this paper.

  \subsection{Implementation of inner boundary condition}\label{sec:Brief overview of the application of UBC in the inner boundary}
\begin{figure}[htpb]
\begin{center}
  \vspace*{0.01\textwidth}
    \includegraphics[width=0.4\linewidth,trim=1 1 1 1, clip]{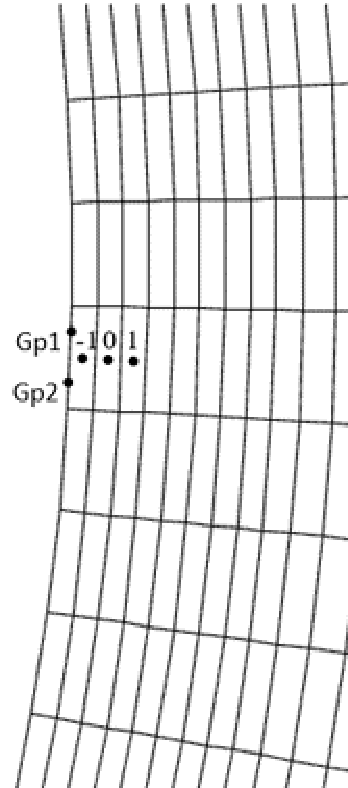}
\end{center}
\caption{2D illustration of the inner boundary cells which lay on the leftmost layer of this picture.}\label{InnerBDcell}
\end{figure}

  Figure~\ref{InnerBDcell} shows a 2D illustration of the inner-boundary cells, for example, cell $-1$, which is situated on the leftmost layer of this picture. Here, Gp1 and Gp2 are two Gaussian points on the inner-boundary face and points $-1$ and $0$ are the cell centroids of cell $-1$ and cell $0$, respectively. To update solutions on cell $0$, we should calculate inviscid flux and heat flux through the surface of cell $0$, including the interface shared by cell $-1$ and cell $0$. It means we should calculate the reconstruction formulation of variables inside cell $-1$ and provide the values of these variables on the interface shared by cell $-1$ and cell $0$ by the formulation of this reconstruction.

  In this paper, we use the reconstruction stencil that includes  the Gaussian points on the inner-boundary face, centroid point of the inner-boundary cell, cell $-1$ for example, and centroid points of the five neighbouring cells of cell $-1,$ which share an interface with this hexahedral cell. Based on the former works \citep{Feng_2007,FengandLiu2019,Feng_2021,Groth2000,Wang_2022}, we implement the condition of no `backflow' at these Gaussian points of the inner-boundary face for both quasi-steady coronal and time-dependent CME simulations. During these simulations, the inner boundary conditions are divided into two cases according to the flow with the radial speed of $v_{r}$ in the cell adjacent to the inner boundary cell in the radial direction. We present the two cases below:

Case 1: If $v_{r}<0$, then
$\frac{\partial\rho}{\partial r}=0$, $\frac{\partial p}{\partial r}=0$, $\frac{\partial T}{\partial r}=0$, $\mathbf{v}=0$ and $\mathbf{B}=\left\{\begin{array}{c}
\mathbf{B}_{0}, \text { for quasi-steady simulation } \\
\mathbf{B}_{0}+\mathbf{B}_{FR},  \text { for CME simulation }
\end{array}\right.$.

Case 2: If $v_{r}\ge 0$, then
$\rho=1$, $p=\frac{1}{\gamma}$, $T=1$, $\frac{\partial v_{r}}{\partial r}=0$, $v_{t1}=0$, $v_{t2}=0$ and $\mathbf{B}=\left\{\begin{array}{c}
\mathbf{B}_{0}, \text { for quasi-steady simulation, } \\
\mathbf{B}_{0}+\mathbf{B}_{FR},  \text { for CME simulation. }
\end{array}\right.$here $v_{t1}$ and $v_{t2}$ are tangential velocities.

   Similarly to \cite{Feng_2021} and \cite{Wang_2022}, these boundary conditions are used to constrain the following reconstruction formulations:
\begin{equation}\label{Eq:FlowfieldIBD}
\begin{aligned}
X_{-1}(\mathbf{x})=&\left.X\right|_{-1}+\psi_{-1}\left.\left(\nabla X\right)\right|_{-1}\cdot\left(\mathbf{x}-\mathbf{x}_{-1}\right), \\ X \in &\{\rho,u,v,w,\mathbf{B}_1,\mathbf{B}_0,p,T\},\
\end{aligned}
\end{equation}
   where $\mathbf{x}$ is a point inside cell $-1$, $\mathbf{x}_{-1}$ is the centroid position of cell $-1$, $\left.X\right|_{-1}$ is the corresponding variable at $\mathbf{x}_{-1}$ and $\left.X\right|_{-1}$ is constant during the simulation, and $\left.\left(\nabla X\right)\right|_{-1}=\left.\left(\frac{\partial X}{\partial x},\frac{\partial X}{\partial y},\frac{\partial X}{\partial z}\right)\right|_{-1}$ denotes the derivative of $X$ at $\mathbf{x}_{-1}$. In this paper, $\left.\left(\nabla X\right)\right|_{-1}$ is obtained by solving a least-square problem \citep{BARTH1991,BARTH1993}, and $\psi_{-1}$ is defined as the Barth limiter \citep{Barth1989}.

  \subsection{ Pseudo-time marching method  }\label{sec: Pseudotimemarching}
   In this paper, we perform a backwards Euler temporal integration on Eq.~(\ref{MHDinsolarwind}) and reach the following equations,
\begin{equation}\label{implicitbackwardEuler}
V_{i}\frac{\Delta \mathbf{U}^n_{i}}{\Delta t}+\mathbf{R}^{n+1}_{i}=\mathbf{0}.\
\end{equation}
   The superscripts $^{`n'}$ and $^{`n+1'}$ denote the time level, $\mathbf{R}^{n+1}_i=\sum\limits_{j=1}^{6}\mathbf{T}_{8}^{-1}\mathbf{f}_{ij}\left(\mathbf{U}^{n+1}_{nL},\mathbf{U}^{n+1}_{nR}\right)\Gamma_{ij}-V_{i}\mathbf{S}^{n+1}_i$, means the residual operator over cell $i$ at the $(n+1)$-th time level, $\Delta \mathbf{U}_i^{n}=\mathbf{U}_i^{n+1}-\mathbf{U}_i^{n}$, is the solution increment between the $n$-th and $(n+1)$-th time level, and $\Delta t = t^{n+1}-t^n$ is the time increment. Unlike explicit methods, the implicit approach has flexibility in selecting a large time step exceeding the CFL condition. In the background coronal simulation, the time variable, $t$, does not refer to a physical time but a relaxation time used to get a quasi-steady state coronal structure. As in \cite{Wang_2022}, this backwards Euler temporal integration efficiently calculates quasi-steady-state coronal structures. More specifically, we call this quasi-steady-state model the quasi-steady-state SIP-IFVM coronal model.

   As for the time-dependent CME simulation, we need to make the implicit algorithm time-accurate. To improve the temporal accuracy for CME simulations, we further introduced a pseudo time $\tau$ to Eq.~(\ref{implicitbackwardEuler}) and updated the solution during each physical time step $\Delta t$ by solving a steady-state problem on $\tau$. Consequently, we achieve the following equation
\begin{equation}\label{implicitbackwardEulerPseudotime}
V_{i}\frac{\Delta \mathbf{U}_{i}}{\Delta \tau}+\left(V_{i}\frac{\Delta \mathbf{U}^n_{i}}{\Delta t}+\mathbf{R}^{n+1}_{i}\right)=\mathbf{0}.\
\end{equation}
   Here, $\Delta \tau$ is a pseudo time step and $\Delta \mathbf{U}_{i}$ is the solution increment during $\Delta \tau$. More specifically, we call this time-accurate model the 'time-dependent SIP-IFVM coronal model'.

   In this paper, we solve Eq.~(\ref{implicitbackwardEulerPseudotime}) by using backwards Euler method, as detailed below:
\begin{equation}\label{backwardEulerPseudotime}
V_{i}\frac{\mathbf{U}_{i}^{n,m+1}-\mathbf{U}_{i}^{n,m}}{\Delta \tau}=\left(V_{i}\frac{\mathbf{U}^n_{i}-\mathbf{U}_i^{n,m+1}}{\Delta t}-\mathbf{R}^{n,m+1}_{i}\right).\
\end{equation}
   In Eq.~(\ref{backwardEulerPseudotime}), $\mathbf{U}_{i}^{n,m+1}$ and $\mathbf{U}_{i}^{n,m}$ denote the solution variables on $\tau^{n,m+1}$ and $\tau^{n,m}$, and $\mathbf{U}_{i}^{n,0}=\mathbf{U}_{i}^{n}$. The superscripts `$^{n,m}$' and `$^{n,m+1}$' denote the corresponding variables on the $m$-th and $(m+1)$-th pseudo time level during the $n$-th physical time step. Similar to \cite{Wang_2022}, we implement an approximate local time linearisation for the residual operator $\mathbf{R}_i^{n,m+1}$ at $\left(\tau^{n,m},\mathbf{U}_i\right)$ with respect to pseudo time as below,
$$\mathbf{R}_i^{n,m+1}\approx\mathbf{R}_i^{n,m}+\left(\frac{\partial{\mathbf{R}^{'}_i}}{\partial{\mathbf{U}_i}}\right)^{n,m}\Delta \mathbf{U}_i^{n,m}+\sum\limits_{j=1}^6\left(\frac{\partial{\mathbf{R}^{'}_i}}{\partial{\mathbf{U}_j}}\right)^{n,m}\Delta \mathbf{U}_j^{n,m},$$
   where $\Delta \mathbf{U}_{i/j}^{n,m}=\mathbf{U}_{i/j}^{n,m+1}-\mathbf{U}_{i/j}^{n,m}$ is the solution increment at the $m$-th pseudo time step of the $n$-th physical time step in cell $i$ and the $j$-th neighboring cell cell $j$, and $\mathbf{R}_i^{'}=\sum\limits_{j=1}^{6}\mathbf{T}_{8}^{-1}\mathbf{f}_{ij}^{'}\left(\mathbf{U}_{nL},\mathbf{U}_{nR}\right)\Gamma_{ij}-V_{i}\mathbf{S}_i$. As in \cite{Wang_2022}, the modified numerical flux $\mathbf{f}_{ij}^{'}\left(\mathbf{U}_{nL},\mathbf{U}_{nR}\right)$ is calculated by adding an appropriate viscous term to $\mathbf{f}_{ij}\left(\mathbf{U}_{nL},\mathbf{U}_{nR}\right)$ \citep{Otero2015,OteroAcc2015} to help maintain a diagonally dominant Jacobian matrix with a smaller condition number.
   Consequently, Eq.~(\ref{backwardEulerPseudotime}) can be written as
\begin{equation*}
\begin{aligned}
\frac{V_i}{\Delta \tau}\left(\mathbf{U}_i^{n,m+1}-\mathbf{U}_i^{n,m}\right)=&\frac{V_i}{\Delta t}\left(\mathbf{U}_i^n-\mathbf{U}_i^{n,m+1}\right)-\mathbf{R}^{n,m}_{i}-\\ &\left(\frac{\partial\mathbf{R}_i^{'}}{\partial\mathbf{U}}\right)^{n,m}\left(\mathbf{U}^{n,m+1}-\mathbf{U}^{n,m}\right).\
\end{aligned}
\end{equation*}
   This means:
\begin{equation}\label{modifiedglobalPseIBElinearized}
\begin{aligned}
&\left(\frac{V_i}{\Delta \tau}\mathbf{I}+\frac{V_i}{\Delta t}\mathbf{I}+\left(\frac{\partial\mathbf{R}_i^{'}}{\partial\mathbf{U}_i}\right)^{n,m}\right)\left(\mathbf{U}_i^{n,m+1}-\mathbf{U}_i^{n,m}\right)=\\&\frac{V_i}{\Delta t}\left(\mathbf{U}_i^n-\mathbf{U}_i^{n,m}\right)-\mathbf{R}_{i}^{n,m}-\left(\frac{\partial\mathbf{R}_i^{'}}{\partial\mathbf{U}_j}\right)^{n,m}\left(\mathbf{U}_j^{n,m+1}-\mathbf{U}_j^{n,m}\right).\
\end{aligned}
\end{equation}
   As a result, we reach the following linearised system
\begin{equation}\label{uniformmodifiedglobalPseIBElinearized}
\mathbf{A}^{''}_{8N\times8N}\left(\Delta \mathbf{U}^{n,m}\right)_{8N}=\mathbf{b}_{8N}
,\end{equation}
   where $\mathbf{A}^{''}_{8N\times8N}=\left(\frac{V}{\Delta \tau}+\frac{V}{\Delta t}+\left(\frac{\partial\mathbf{R}^{'}}{\partial\mathbf{U}}\right)^{n,m}\right)_{8N\times8N}$, $\left(\Delta \mathbf{U}^{n,m}\right)_{8N}=\left(\mathbf{U}^{n,m+1}\right)_{8N}-\left(\mathbf{U}^{n,m}\right)_{8N}$ and $\mathbf{b}_{8N}=\frac{V}{\Delta t}\left(\left(\mathbf{U}^n\right)_{8N}-\left(\mathbf{U}^{n,m}\right)_{8N}\right)-\left(\mathbf{R}^{n,m}\right)_{8N}$ with $N$ denoting the number of cells in computational domain.

   In this paper, we solve Eq.~(\ref{uniformmodifiedglobalPseIBElinearized}) by the parallel implicit LU-SGS method \citep{Feng_2021,Wang_2022}. In Eq.~(\ref{uniformmodifiedglobalPseIBElinearized}), $\Delta t$ is gradually increased to $\chi \cdot \tau_{flow}$ where $\chi$ is an adjustable parameter and $\tau_{flow}$ is a reference time length that is the same as defined in \cite{Feng_2021} and \cite{Wang_2022}. In our simulations, the values of $\tau_{flow}$ are $0.112$ and $0.057$ hours for CR 2219 and the ad hoc case with initial magnetic field strength multiplied by 5, respectively.
   Afterwards, $\Delta t=\chi \cdot \tau_{flow}$ advances solutions on the following physical time steps. The time-step size, $\Delta t$, can generally affect the solution accuracy and computational efficiency for time-dependent simulations. Selecting a considerable time step in the implicit method usually leads to a loss in temporal accuracy. In contrast, a small time step leads to more time steps and requires more computing resources. To find a suitable time step that can both maintain the required temporal accuracy and desired high computational efficiency for CME simulations, we set $\chi=1,~0.5,~0.25,~0.125,$ respectively. We then compared the effects of different $\Delta t$ on simulation results in Sect. \ref{sec:Numerical Results}.

   Also, the pseudo-time-step size, $\Delta \tau$, can affect the convergence rate of the steady-state simulation in pseudo-time $\tau$. We set $\Delta \tau=10^{20}$ in the initial pseudo time step of the $n$-th physical time step and get $\left(\mathbf{U}^{n,1}\right)_{8N}=\left(\mathbf{U}^{n,0}\right)_{8N}+\left(\Delta \mathbf{U}^{n,0}\right)_{8N}$. By this mean, $\left(\mathbf{U}^{n,1}\right)_{8N}$ serves as a good preliminary guess for the steady-state solution on $\tau$. In the following pseudo time steps during the $n$-th physical time step, we set $\Delta \tau=\Delta t$ to gradually evolve the solution from $\left(\mathbf{U}^{n,1}\right)_{8N}$ to the steady state solution on $\tau$ of the $n$-th physical time step. This strategy helps guarantee both the numerical stability and computational efficiency. Meanwhile, the simulation is judged to reach the steady state condition on $\tau$ of the $n$-th physical time step when $\frac{\left|\left(\Delta \mathbf{U}^{n,m}\right)_{8N}\right|}{N}<\epsilon_1$ or $\frac{\left|\left(\Delta \mathbf{U}^{n,m}\right)_{8N}\right|}{\left|\left(\Delta \mathbf{U}^{n,0}\right)_{8N}\right|}<\epsilon_2$. Here, $\epsilon_1$ and $\epsilon_2$ are two adjustable small parameters and we set $\epsilon_1=10^{-5}$ and $\epsilon_2=\frac{1}{500}$ in this paper. Eventually, we set $\left(\mathbf{U}^{n+1}\right)_{8N}=\left(\mathbf{U}^{n,m+1}\right)_{8N}$ and stop the pseudo-time simulation of the $n$-th physical time step when the steady-state condition is satisfied at the $m$-th pseudo-time step of the $n$-th physical time step. For better computational efficiency, we limit the number of pseudo-time steps during a physical time step to be no more than $N_{\tau}$ and set $\left(\mathbf{U}^{n+1}\right)_{8N}=\left(\mathbf{U}^{n,{N_{\tau}}}\right)_{8N}$ once the number of pseudo-time steps during a physical time step reaches $N_{\tau}$. In this paper, we set $N_{\tau}=5$.

   Considering that the forwards-backwards sweep of an LU-SGS iteration can only update solutions of inner cells of a processor, but not solutions of ghost cells. The solution information in ghost cells is also required in parallel LU-SGS method \citep{Feng_2021} and so, we need to carefully perform synchronised MPI data communication between different processors in the parallel LU-SGS method \citep{OteroAcc2015,Petrov2017,Sharov2000} to avoid a degradation of the convergence rate. We refer to Appendix \ref{sec:data communication between different components} for a description of how we implemented the data communication in our six-component composite grid system.

\section{Numerical results}\label{sec:Numerical Results}
   In this step, we first used the quasi-steady-state SIP-IFVM coronal model to mimic the quasi-steady-state solar corona of CR 2219. Subsequently, the time-dependent SIP-IFVM coronal model was applied to investigate the CME evolution and propagation procedure in the background solar corona. The quasi-steady-state SIP-IFVM can be seen as a specific case of the time-dependent SIP-IFVM coronal model, consisting of only one pseudo-time iteration.
   
   CR 2219 is around the solar minimum of the solar cycle (SC) 24 and continues from June 29, 2019 to July 26, 2019. The synoptic map\footnote{\url{https://gong.nso.edu/adapt/maps/gong/}} of the radial photospheric magnetic field centred on 2019 July 2 was used as the inner-boundary condition for the magnetic field during the quasi-steady-state coronal simulation. Following this quasi-steady-state simulation, an RBSL flux rope with a theoretical-based `S-shaped' axis path is inserted into the steady corona to trigger the CME event. We compare the CME simulation results using different large time-step sizes ($\rm CFL \gg 1$) with those, using small time-step sizes ($\rm CFL=1$) to validate the model's capability to deliver accurate and efficient time-varying simulations. 
   
   Furthermore, to demonstrate the implicit MHD model's capability of dealing with low plasma $\beta$ problems, we performed an ad hoc simulation by artificially enhancing the initial magnetic field of the quasi-steady-state coronal simulation and CME simulation by factors of 5 and 2.5, respectively, while keeping the other initial parameters unchanged. This led to a low plasma $\beta$ of about $5 \times 10^{-4}$ and a strong magnetic field strength of about 53 $\rm{Gauss}$ for the background corona, while the magnetic field strength of the flux rope reaches 34 $\rm{Gauss}$ near the footpoints of the flux rope's `S- shaped' axis path. In addition, we carried out an Orszag-Tang MHD vortex simulation in Appendix A to show that the novel pseudo-time-marching method and parallel LU-SGS method adopted in this paper is capable of simulating small-scale unsteady-state flows. 

   In this work, all the calculations were performed on the Tier-2 supercomputer of Vlaams Supercomputer Centrum\footnote{\url{https://www.vscentrum.be/}}. Both simulations were completed on 192 CPU cores. The quasi-steady state simulations calculated by the quasi-steady-state SIP-IFVM coronal model for CR 2219 and the ad hoc simulation reach the steady state after 776 and 1243 iterations, while the wall-clock time durations are $0.12$ and $0.18$ hours, respectively. Moreover, the wall-clock times are less than $0.7$ hours for the time-dependent CME simulations of six hours of physical time when a large time step $\Delta t \geq 0.125 \cdot \tau_{flow}$ was adopted for both simulations. Here, $\tau_{flow}$ is the predefined reference time length mentioned in Sect.~\ref{sec: Pseudotimemarching} that we used to constrain the time-step sizes. In comparison, in the CME simulations with a small time step with $\rm CFL=1$, the corresponding wall-clock times are $2.61$ and $18.03$ hours for CR 2219 and the ad hoc simulation, respectively. It demonstrates that the SIP-IFVM MHD coronal model is very efficient and numerically stable in the computations of both steady-state background coronal structures and time-dependent CME events.

   In the rest of this section (and Appendix A), we present the results of the quasi-steady coronal simulations of CR 2219, the ad hoc case with artificially enlarged magnetic field, the corresponding time-varying CME simulations, and the small-scale Orszag-Tang MHD vortex simulations.

  \subsection{Background coronal simulation}\label{sec:Quasisteadystatesimulation}
   This subsection presents the steady-state simulation results for CR 2219. We compare the results with the solar coronal observations and compare the simulation results calculated by this quasi-steady-state SIP-IFVM coronal model and by the SIP-IFVM coronal model's explicit counterpart. We adopt
   the following explicit second-order Runge-Kutta scheme (ERK2) in Eq.~(\ref{2orderRK}) in this explicit MHD model and call it SIP-EFVM coronal model, namely, 

\begin{equation}\label{2orderRK}
\begin{aligned}
&\mathbf{U}^{(1)}=\mathbf{U}^{n}+\Delta t \mathbf{R}\left(\mathbf{U}^n\right)\\
&\mathbf{U}^{n+1}=\frac{1}{2}\mathbf{U}^{n}+\frac{1}{2}\left(\mathbf{U}^{(1)}+\Delta t \mathbf{R}\left(\mathbf{U}^{(1)}\right)\right).
\end{aligned}
\end{equation}
Here, the formula for calculating $\Delta t$ in Eq.~(\ref{2orderRK}) is the same as in Eq.~(\ref{implicitbackwardEuler}) and we set $\rm CFL=0.5$.

   In the quasi-steady state coronal simulation calculated by the explicit coronal model, the steady-state condition was reached after 78767 iterations with a time step of around $5.6\times10^{-4}$ hours for CR 2219, while the wall-clock time was 8.25 hours. In addition, the average relative difference in proton number density, ${\rm RD}_{{\rm ave},\rho}$, and radial velocity, ${\rm RD}_{{\rm ave},{V_r}}$, between the steady-state results simulated by the quasi-steady-state SIP-IFVM and SIP-EFVM coronal models was only $3.05\%$ and $3.43\%$. Here, ${\rm RD}_{{\rm ave},\rho}=\sum\limits_{i=1}^N\big|\rho_i^{\rm SIP-IFVM}-\rho_i^{\rm SIP-EFVM}\big|\big/\sum\limits_{i=1}^N\rho_i^{\rm SIP-EFVM}$, and ${\rm RD}_{{\rm ave},{V_r}}=\sum\limits_{i=1}^N\big|{V_r}_i^{\rm SIP-IFVM}-{V_r}_i^{\rm SIP-EFVM}\big|\big/\sum\limits_{i=1}^N {V_r}_i^{\rm SIP-EFVM}$, while the superscripts `$^{\rm SIP-IFVM}$' and `$^{\rm SIP-EFVM}$' denote the corresponding variable calculated by the SIP-IFVM and SIP-EFVM, respectively, and $N$ is the number of cells in the computational domain. We also conducted a quasi-steady-state coronal simulation by the time-dependent SIP-IFVM coronal model. The steady-state condition was reached after 515 time steps with a wall-clock time of 0.39 hours. The average relative difference in proton number density and radial velocity simulated by the time-dependent SIP-IFVM and SIP-EFVM coronal models are only $2.54\%$ and $1.93\%$, respectively. 
   
   This means that our quasi-steady-state and time-dependent SIP-IFVM coronal models gain a speeding up of $68.7\times$ and $21.2\times$ for CR 2219, compared to the explicit coronal model, while maintaining consistency in the steady-state coronal structures. In the following, we demonstrate the simulation results calculated by the quasi-steady-state SIP-IFVM coronal model.

  \subsubsection{The open-field regions in the solar corona}\label{sec:Distributions of the open-field regions}
   Coronal holes (CHs) are dark regions in the images observed in extreme ultraviolet (EUV) and soft X-ray channels due to low plasma density caused by the magnetic field lines from CHs that are open to interplanetary space. Overall, CHs are the most prominent features in the solar corona because their distributions vary from different solar activity phases \citep{FengMa2015,Feng_2017,FengandLiu2019,Frazin2007,Hayes2001,Linker1999JGR,Petrie2011SoPh}. Three types of CHs can be identified in the EUV and soft X-ray images of the solar corona. Polar CHs are located at both solar poles and often stretch to low latitudes, sometimes across the solar equator. Isolated CHs, often seen near solar maxima, are detached from polar CHs and scatter at low and middle latitudes. Transient CHs are associated with solar eruptive events, such as coronal mass ejections, solar flares, and eruptive prominences.

\begin{figure*}[htpb!]
\begin{center}
  \vspace*{0.01\textwidth}
    \includegraphics[width=0.8\linewidth,trim=1 1 1 1, clip]{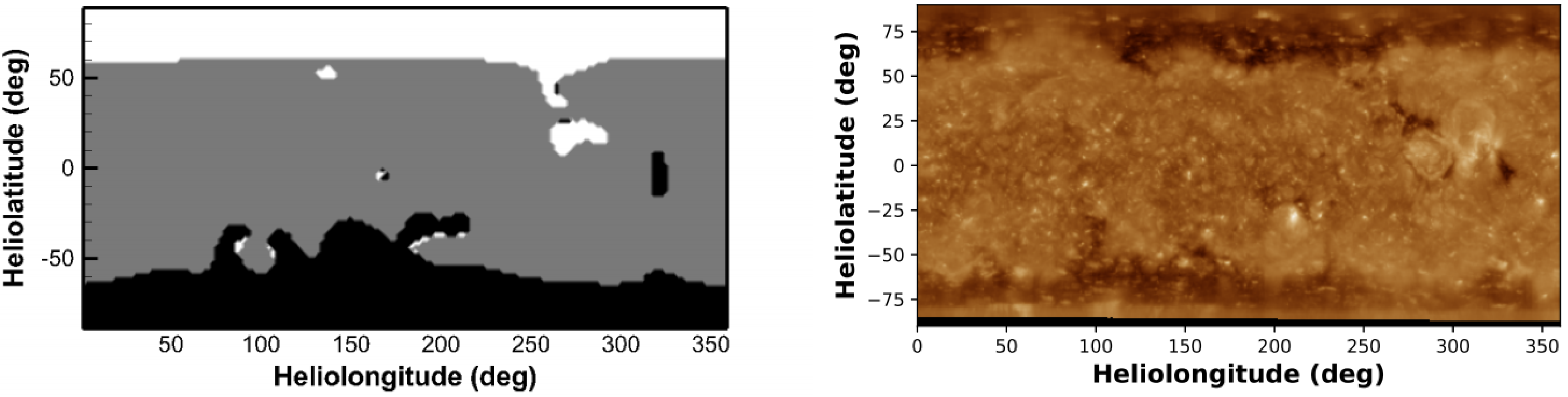}
\end{center}
\caption{Synoptic maps of the open-field regions modelled by the quasi-steady-state SIP-IFVM MHD model (left), and extreme ultraviolet observations from the 193 {\AA} channel of AIA on board SDO (right) for CR 2219. In the synoptic map, the white and black patches denote open-field regions where the magnetic field lines point outwards and inwards to the Sun, respectively, and the grain patches denote the close-field regions.}\label{Cr2219Cr2111CHatsolarsurface}
\end{figure*}Figure~\ref{Cr2219Cr2111CHatsolarsurface} illustrates synoptic maps of the observations (right) from the Atmospheric Imaging Assembly (AIA) telescope on board the Solar Dynamics Observatory \citep{Lemen2012}\footnote{\url{https://sdo.gsfc.nasa.gov/data/synoptic/}}, and the distributions of open- and closed-magnetic field regions achieved from the simulations (left) for CR 2219. The synoptic maps of observation are generated by concatenating a series of meridian strips taken from full-disk images in a time duration of a complete CR \citep{Hamada2018}. The synoptic maps of these observations and simulations reveal that the simulation roughly captures the polar and isolates coronal holes. 
   
   This simulation aptly reproduces the northern polar CH covering almost all longitudes except the patch between $30^{\circ}$ and $120^{\circ}$ with the latitudes of $60^{\circ} {\rm N}$ pole-wards. It also well captures the southern polar CH, almost spanning all longitudes for latitudes of $65^{\circ} {\rm S}$ pole-wards. In both simulation and observation results, the southern polar CHs between longitudes of $80^{\circ}$ and $180^{\circ}$ extend from $65^{\circ} {\rm S}$ to about $30^{\circ} {\rm S}$ and the northern polar CH extend towards the equator from $(\theta_{\rm lat},\phi_{\rm long})=(65^{\circ} {\rm N}, 265^{\circ})$ and reach an isolated CH centred at $(\theta_{\rm lat},\phi_{\rm long})=(15^{\circ} {\rm N}, 280^{\circ})$, where `$\theta_{\rm lat}$' stands for heliographic latitude and `$\phi_{\rm long}$' Carrington longitude. In addition, the isolated CH centered at $(\theta_{\rm lat},\phi_{\rm long})=(-15^{\circ} {\rm S}, 325^{\circ})$ can also be reproduced by the quasi-steady-state SIP-IFVM model. However, the CH extending from the south pole to the solar equator is larger in the simulation results than in the observation results. This may be due to the spacecraft of SDO nearly orbiting in the plane of the solar equator, thereby resulting in the poor observation of polar CHs. The discrepancy between the modelled and observed results in polar regions may also be attributed to inaccurate observations for both polar photospheric magnetic fields, the utilisation of periodic conditions in the longitudinal direction during the simulations, and the coronal evolution during this period. According to both past simulated and observational studies \citep{Abramenko2010,Sun2011,Yang2011}, stronger magnetic fields in polar regions tend to result in larger areas of polar coronal holes, less of a presence in terms  of low- and middle-latitudinal isolated coronal holes, and flatter coronal magnetic neutral lines. In addition, these differences in the solar corona can propagate outwards and cause different manifestations in the heliosphere \citep{RILEY20121}. Therefore, the uncertainty caused by periodically missing and high noise levels presented in the observations of solar polar fields is a critical factor that causes differences between the observations and the results of 3D solar wind MHD simulations.

  \subsubsection{Simulated steady-state solar corona near the Sun}\label{sec:Simulated results near the Sun}
\begin{figure*}[htpb!]
\begin{center}
  \vspace*{0.01\textwidth}
    \includegraphics[width=0.8\linewidth,trim=1 1 1 1, clip]{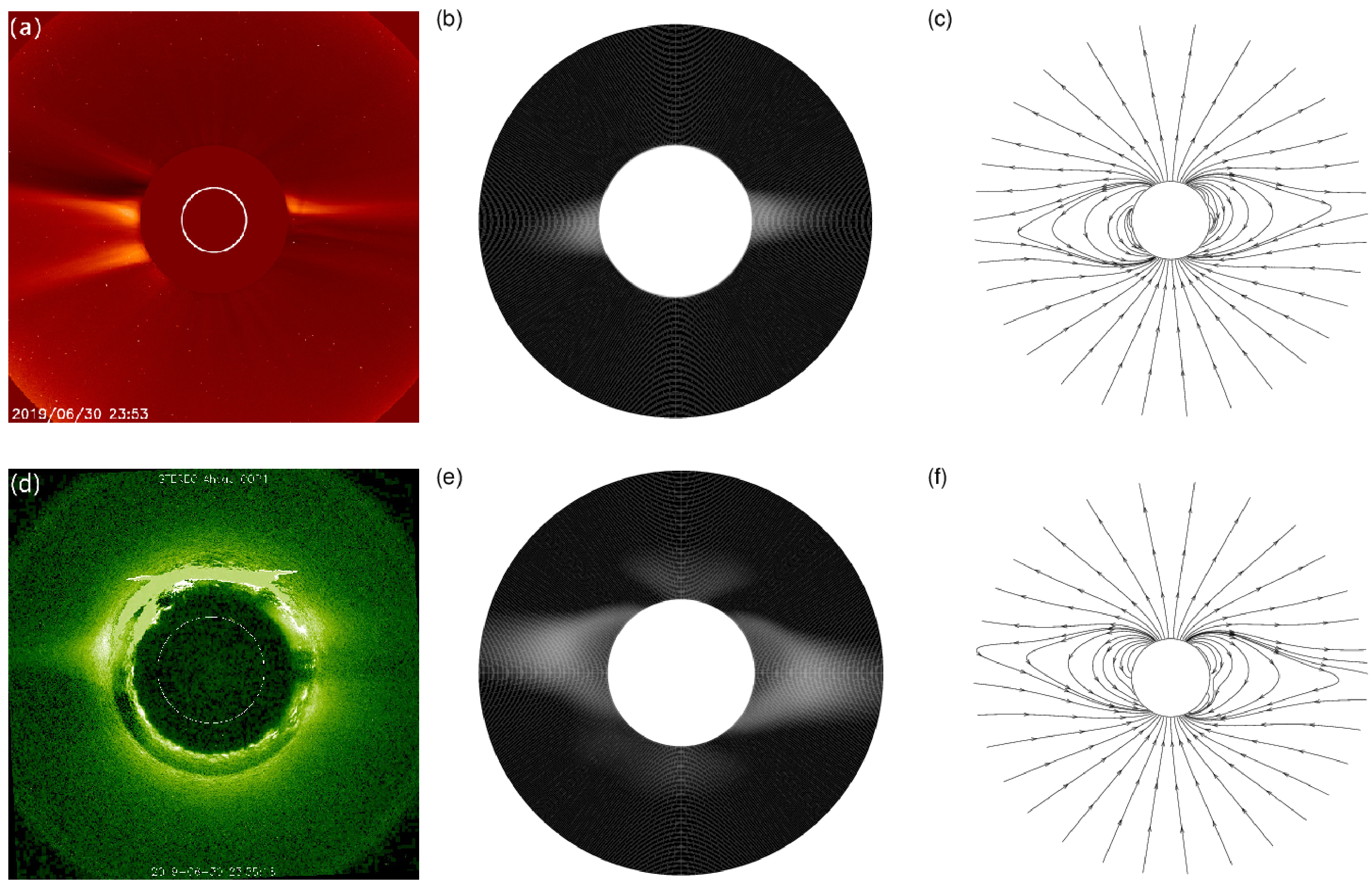}
\end{center}
\caption{White-light pB images observed from LASCO C2/SOHO (a) and COR1/STEREO-A (d) on June 30, 2019, corresponding pB images synthesised from simulation results (b, e), these synthesised images range from 2.3 to $6 R_s$ on the meridian plane of $\phi_{\rm long}=250^{\circ}-70^{\circ}$ (b) and range from 1.4 to $4 R_s$ on the meridian plane of $\phi_{\rm long}=160^{\circ}-340^{\circ}$ (e), respectively, and 2D simulated magnetic field lines from 1 to $5 R_s$ on these two selected meridian planes (c, f).}\label{Cr2177Cr2212PbSky}
\end{figure*}
   The white-light polarised brightness (pB) images can manifest the coronal structures seen at both limbs. In these pB images, bright regions represent high-density coronal structures, such as bipolar streamers and pseudo-streamers. In contrast, dark regions denote low-density coronal structures such as coronal holes \citep{FengMa2015,Feng_2017,FengandLiu2019,Feng2020book,Frazin2007,Hayes2001,Linker1999JGR,Petrie2011SoPh}. Bipolar streamers separate CHs of opposite magnetic polarities while pseudo-streamers separate CHs of the same polarity. In addition, bipolar streamers extend outwards several solar radii from the Sun and they are drawn into a cusp-like structure with a current sheet formed above the helmet streamer \citep{Abbo2015,Feng_2017,FengandLiu2019,Riley2011,Wang_2007}.

   In Fig.~\ref{Cr2177Cr2212PbSky}, we compare white-light pB images from $2.3$ to $6 R_s$ that observed from the Large Angle and Spectrometric Coronagraph C2 \citep{Brueckner1995} on board the Solar and Heliospheric Observatory (SOHO), from  $1.4$ to $4 R_s$, which were  observed by the innermost coronagraph of the Sun-Earth Connection Coronal and Heliospheric Investigation (SECCHI) instrument suite on board the Solar Terrestrial Relations Observatory Ahead (STEREO-A) spacecraft \citep{Howard2008,Kaiser2008TheSM,Thompson2003COR1IC,Thompson2008}, both available at \url{https://stereo-ssc.nascom.nasa.gov/browse/} and synthesised from the results of the SIP-IFVM MHD coronal model (b, e). The observed and modelled images show a bright structure almost horizontally on the west limb in the LASCO-C2 view and the east limb in the STEREO-A view. In the LASCO-C2 and STEREO-A views, two narrow bright structures exist in the observed images, but only one is centred in the simulated result. However, the  centre positions and widths of the bright structures are consistent in both observed and modelled images. From the close-ups of magnetic field lines ranging from 1 to $5 R_s$ on the two selected meridian planes (c, f), we can deduce that the bipolar streamers produce bright structures.

  \subsection{Time-dependent CME simulations}\label{sec:CMEsimulation}
   In this subsection, we describe how we inserted the magnetic field of the RBSL flux rope with a theoretical `S-shaped' axis path to the quasi-steady state solar corona of CR 2219 to trigger CME events. The pseudo-time marching method described in Sect. \ref{sec: Pseudotimemarching} was used to mimic these CME evolution and propagation procedures from the solar surface to around 0.1 AU. 
   First, we performed a CME simulation in the background corona of CR 2219 with a small physical time step, $\Delta t$. Then, we carried out four CME simulations with considerable physical time steps. Next, we compared the simulation results calculated by adopting different large $\Delta t$ and those calculated by adopting small $\Delta t$. We set $\rm CFL=1$ in the simulation, adopting small time step, and constrained $\Delta t \leq \chi \cdot \tau_{flow}$ as described in Sect. \ref{sec: Pseudotimemarching}. Finally, we set $\chi=1,~0.5,~0.25,  ~0.125,$ respectively, in these four CME simulations adopting large values of $\Delta t$.

\begin{figure*}[htpb!]
\begin{center}
     \includegraphics[width=0.7\linewidth,trim=1 1 1 1, clip]{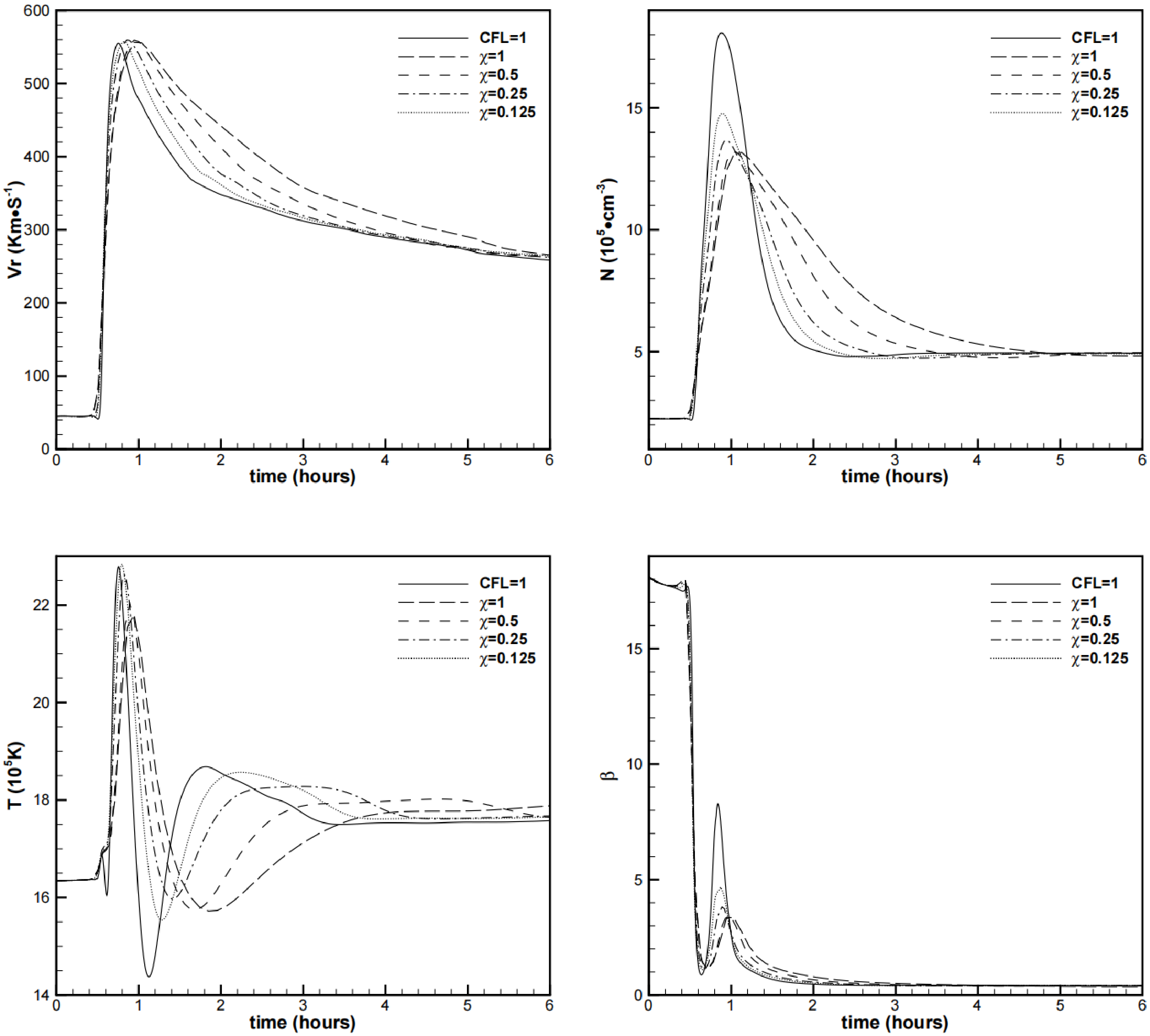}
\end{center}
  \caption{In situ measurements of simulated radial velocity, $v_r~ \left(\rm km~ s^{-1}\right),  $ on the top-left, proton number density $\left(10^5~ \rm cm^{-3}\right)$ on the top-right, temperature $\left(10^5~ \rm K\right)$ on the bottom-left, and plasma $\beta$ on the bottom-right, taken by the virtual satellite placed at $\left(r, \theta, \phi\right)=\left(3 R_s, 0^{\circ}, 250^{\circ}\right)$.}\label{VirtualS1}
\end{figure*}

   As  in \cite{Linan_2023} and \cite{guo2023}, we placed a virtual satellite to monitor the variation pattern of solar corona when disturbed by CMEs. In these CME simulations with large and small time steps, a virtual satellite is placed at point $\left(r, \theta, \phi\right)=\left(3 R_s, 0^{\circ}, 250^{\circ}\right)$ to observe the changes of radial velocity, $V_r$, plasma density, $\rho$, thermal temperature, $T$, and plasma, $\beta$. As illustrated in Fig.~\ref{VirtualS1}, there is a fluctuation of $V_r$, $\rho$, $T$, and $\beta$ at about 0.6 hours of the CME simulations. Afterwards, a peak appears in the profile of $V_r$, $\rho$, and $T$, and a trough appears at the profile of $\beta$. During the period of 0.6 and 1.0 hours of these CME simulations with different time-step sizes, the radial velocities all increase from $50~ \rm km~ s^{-1}$ to about $560~ \rm km~ s^{-1}$, the number densities increase from $2.3 \times 10^5~ \rm cm^{-3}$ to $18 \times 10^5 ~ \rm cm^{-3}$, $14.8 \times 10^5~ \rm cm^{-3}$, $13.7 \times 10^5~ \rm cm^{-3}$, $13.2 \times 10^5~ \rm cm^{-3}$, and $13.2 \times 10^5~ \rm cm^{-3}$ for $\rm CFL=1$ and $\chi=0.125,~0.25,~ 0.5,~1$, respectively. Although there was a delay in time for these parameters at large time steps compared to the results of small time steps, it is less than 0.1 hours for $\chi=0.125$.  Afterwards, the radial velocity and number density decrease to $265~ \rm km~ s^{-1}$ and $4.9 \times 10^5~ \rm cm^{-3}$, respectively. During the period between 0.45 and 0.8 hours, the temperatures increase from $16.35 \times 10^5~ \rm K$ to $22.78 \times 10^5~ \rm K$ and $22.84 \times 10^5 ~\rm K$; then decrease to $14.4 \times 10^5~ \rm K$ and $15.5 \times 10^5~ \rm K$ at 1.12 and 1.25 hours; and then increase to $18.7 \times 10^5~ \rm K$ and $18.5 \times 10^5~ \rm K$ at 1.8 and 2.1 hours; and decrease to $17.6 \times 10^5~ \rm K$ for $\rm CFL=1$ and $\chi=0.125$, respectively. Although there are some differences in the value and arrival time of the peaks and troughs for the temperature profiles, the relative differences are still minimal for $\rm CFL=1$ and $\chi=0.125$. As for  $\beta$, it decreases from about 18 to about 1 during the period of 0.45 and 0.65 hours, then increases to 8.3 and 4.7 at 0.85 hours, and  decreases to 0.4 in the following time for $\rm CFL=1$ and $\chi=0.125$, respectively. It shows that all of these CME simulations with large time steps (especially the large time steps with $\chi=0.125$) exhibit consistent patterns compared with the simulation calculated by small time steps, while the CMEs modelled by different time-step sizes take almost the same physical time to arrive at this virtual satellite. 

   In Table \ref{largeDtVSsmallDtconsieq1}, we further list the average relative differences in proton number density, ${\rm RD}_{{\rm ave},\rho}^{\chi}$, and radial velocity, ${\rm RD}_{{\rm ave},{V_r}}^{\chi}$, between the CME simulation results of large, $\Delta t=\chi \cdot \tau_{flow}$, and small, $\rm CFL =1$, physical time steps at different moments. Here, ${\rm RD}_{{\rm ave},\rho}^{\chi}=\sum\limits_{i=1}^N\big|\rho_i^{\rm \chi}-\rho_i^{\rm CFL=1}\big|\big/\sum\limits_{i=1}^N\rho_i^{\rm CFL=1}$ and ${\rm RD}_{{\rm ave},{V_r}}^{\chi}=\sum\limits_{i=1}^N\big|{V_r}_i^{\chi}-{V_r}_i^{\rm CFL=1}\big|\big/\sum\limits_{i=1}^N {V_r}_i^{\rm CFL=1}$. The superscripts `$^{\chi}$' and `$^{\rm CFL=1}$' denote the corresponding variable calculated at large ($\Delta t=\chi \cdot \tau_{flow}$) and small ($\rm CFL =1$) physical time steps, respectively, and $N$ is the number of cells in the computational domain.
\begin{table*}
\caption{Comparison of CME simulations for 6 hrs of physical time $t$. }
\label{largeDtVSsmallDtconsieq1}
\resizebox{\linewidth}{!}{
\begin{tabular}{lllll}
\hline\noalign{\smallskip}
 Parameters  &   $\chi=1$  & $\chi=0.5$ & $\chi=0.25$ & $\chi=0.125$\\
\noalign{\smallskip}\hline\noalign{\smallskip}
 wall-clock time (hours) & 0.08 & 0.14 & 0.26 & 0.43\\
 ${\rm RD}_{{\rm ave},\rho}^{\chi}$ \& ${\rm RD}_{{\rm ave},{V_r}}^{\chi}$ at $t$=1hr & $2.19\%$~\&~$1.54\%$ & $1.75\%$~\&~$1.16\%$ & $1.17\%$~\&~$0.72\%$
 & $0.70\%$~\&~$0.40\%$\\
 ${\rm RD}_{{\rm ave},\rho}^{\chi}$ \& ${\rm RD}_{{\rm ave},{V_r}}^{\chi}$ at $t$=3hrs & $3.23\%$~\&~$4.68\%$ & $2.24\%$~\&~$2.86\%$ & $1.59\%$~\&~$1.49\%$ &$1.14\%$~\&~$0.71\%$\\
 ${\rm RD}_{{\rm ave},\rho}^{\chi}$ \& ${\rm RD}_{{\rm ave},{V_r}}^{\chi}$ at $t$=5hrs & $3.52\%$~\&~$4.29\%$ & $2.66\%$~\&~$2.33\%$ & $2.09\%$~\&~$1.18\%$ &$1.62\%$~\&~$0.64\%$\\
\noalign{\smallskip}\hline
\end{tabular}
}
\end{table*}

   It can be seen that the average relative differences of both density and radial velocity decrease with the reduction of physical time-step size. The relative differences in the density and radial velocity are below $3\%$ at different moments with $\chi\leq0.5$. This indicates that the relative differences in CME simulation results obtained using large and small time steps in the time-dependent SIP-IFVM model are no greater than the relative differences between steady-state simulation results computed by the quasi-steady-state SIP-IFVM and SIP-EFVM, while offering significantly higher computational efficiency. Considering that the computational time of 6 hours of physical time is only about 0.43 hours when $\chi=0.125$, the computation efficiency can still be very high when adopting a smaller $\chi$ to get more accurate simulation results at the expense of an acceptable reduction in computation efficiency. We can adjust the physical time-step sizes according to the temporal accuracy required for our specific research or practical application work.
We demonstrate some CME simulation results calculated with $\chi=0.125$ in the following.

\begin{figure*}[htpb!]
\begin{center}
     \includegraphics[width=0.7\linewidth,trim=1 1 1 1, clip]{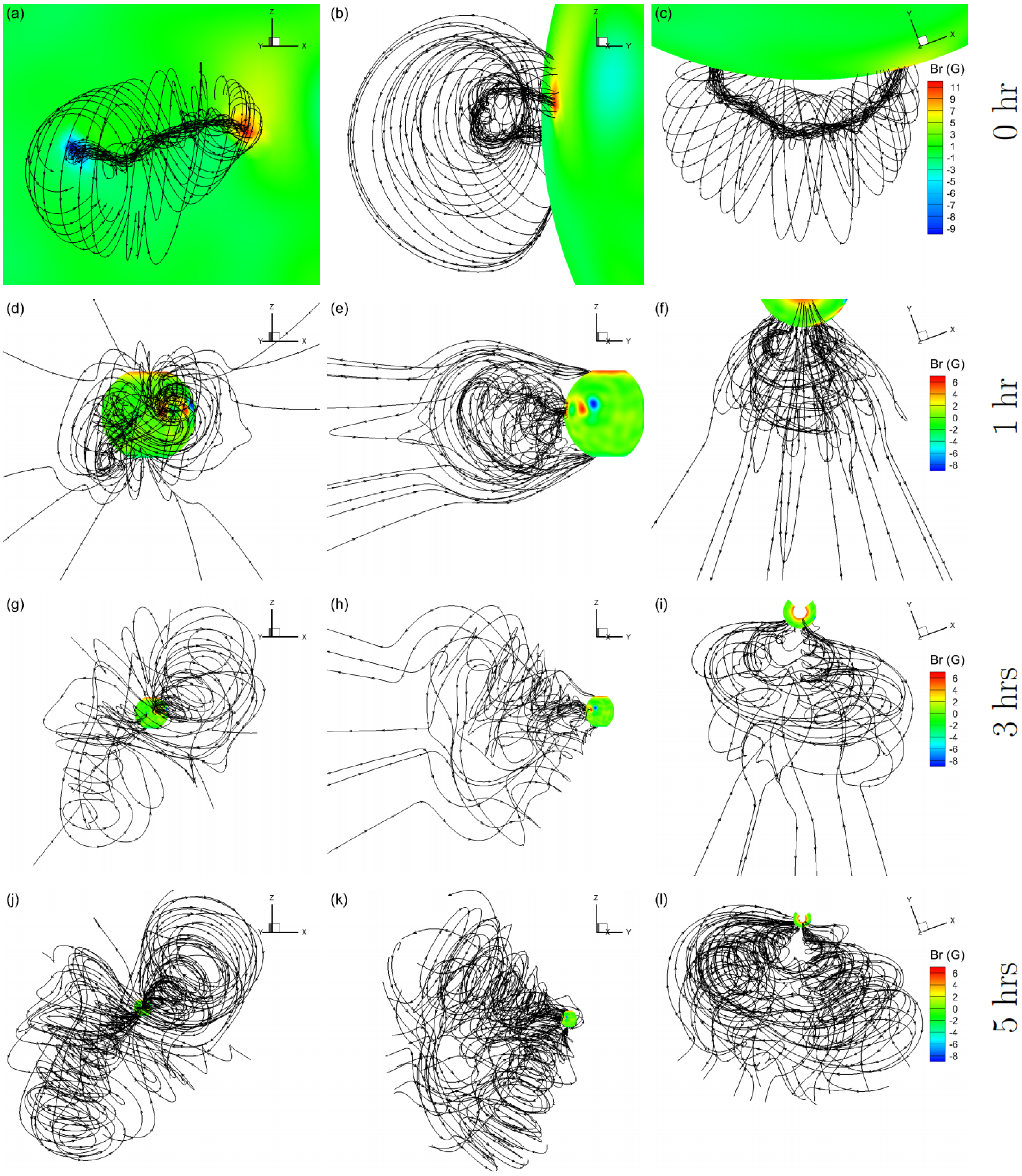}
\end{center}
  \caption{3D view of the magnetic field topology in the corona of CR 2219. These solid lines are representative magnetic field lines displaying the global evolution of the CME and traced from magnetic field in the region of $\left(1R_s \leq r \leq 20 R_s\right)\times\left(22.5^{\circ} \leq \theta \leq 157.5^{\circ}\right) \times \left(29^{\circ} \leq \phi \leq 299^{\circ}\right)$ which encloses the theoretical `S' shape flux rope. Rows 1-4 correspond to the simulation results of the CME simulation at 0, 1, 3 and 5 hours, respectively. The magnetic field lines illustrated in the left panel (a, d, g, j) are viewed from a direction of $\left(\theta,\phi\right)=\left(90^{\circ},250^{\circ}\right)$,  middle panel (b, e, h, k) from a direction of $\left(\theta,\phi\right)=\left(90^{\circ},340^{\circ}\right),$ and  right panel (c, f, i, l) from the direction of the Z axis, these three directions of sight are orthogonal with respect to each other.}\label{CMEsmagneticfileld3DCR2219}
\end{figure*}
   In Fig.~\ref{CMEsmagneticfileld3DCR2219}, we present snapshots of the magnetic field lines at 0, 1, 3, and 5 hours to demonstrate the propagation of the theoretical `S-shaped' flux rope in background coronal structures of CR 2219. These magnetic field lines are traced from the CME simulation results in a region of $\left(1R_s \leq r \leq 20 R_s\right)\times\left(22.5^{\circ} \leq \theta \leq 157.5^{\circ}\right) \times \left(29^{\circ} \leq \phi \leq 299^{\circ}\right)$ and effectively capture the significant changes in the overall morphology of the CME flux rope as it propagates outwards. The magnetic field lines are viewed in three orthogonal directions in the left, middle, and right panels. The left panels are viewed in the direction of $\left(\theta,\phi\right)=\left(90^{\circ},250^{\circ}\right)$, and the sight directions are perpendicular to the meridian, which is parallel to the line connected by the flux rope's two footpoints. The middle panels are obtained by rotating the left panel $90^{\circ}$ clockwise along the $Z$ axis, and the right panels are obtained by rotating the left panels $90^{\circ}$ anticlockwise along the radial direction, which is parallel to the sight directions in the middle panels. It can be seen that the volume overlaid with the CME flux rope expands gradually, which may be attributed to the magnetic-pressure gradient between the flux rope and the surrounding solar atmosphere \citep{Scolini2019}, as well as the magnetic reconnection occurring between the legs of the overlying field lines \citep{Guo2023b}. Also, the topology of the magnetic field lines of the CME flux rope reveals a consistent evolution pattern with those simulated by the poly tropic MHD model \citep{guo2023}, but with a faster-expanding velocity and more realistic thermodynamic evolution. 
   
\begin{figure*}[htpb!]
\begin{center}
    \includegraphics[width=0.6\linewidth,trim=1 1 1 1, clip]{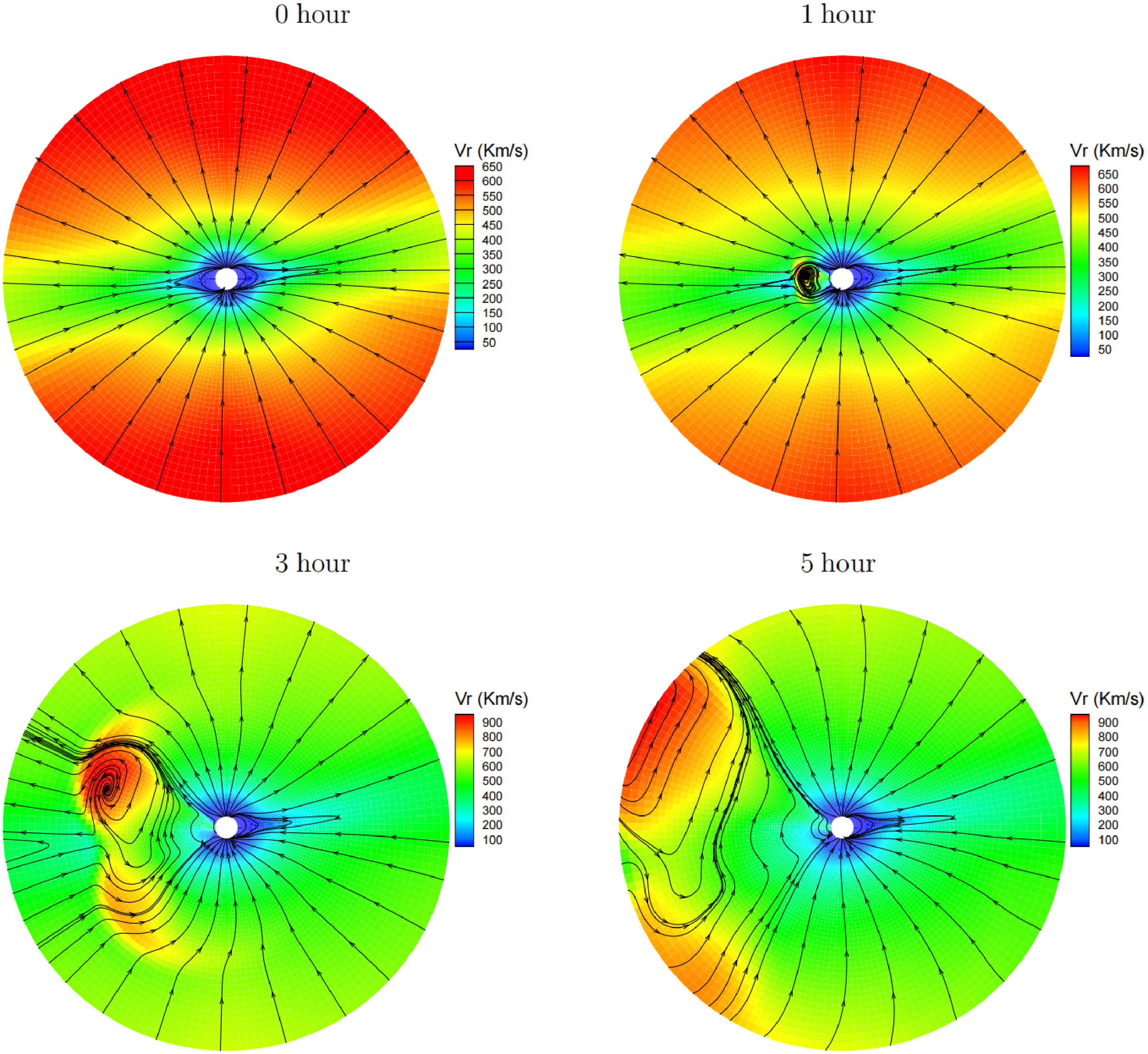}
\end{center}
  \caption{Magnetic field lines from 1 to $20 R_s$ overlaid on contours of the radial velocity on the meridian planes of $\phi_{\rm long}=250^{\circ}-70^{\circ}$ for CR 2219. The left, middle, and right panels correspond to the simulation results at 0, 1, 3, and 5 hours of the CME simulation, respectively.}\label{CMEsVratMeridians}
\end{figure*}

   Furthermore, we present snapshots of the radial speed $V_r$ for CR 2219 at 0 (left-top), 1 (right-top), 3 (left-bottom), and 5 (right-bottom) hours of the CME propagation process in Fig.~\ref{CMEsVratMeridians}. These 2D modelled contours of radial velocity are superimposed with magnetic field lines and range from 1 to $20 R_s$ on the meridian of $\phi_{\rm long}=250^{\circ}-70^{\circ}$. It can be seen that the exceptionally high speed appears at the regions where magnetic field lines change sharply, and the radial velocity can reach $900~ \rm km~ s^{-1}$, which is consistent with the range of the speeds of observed CMEs \citep{Chen2011}. It demonstrates that the model reproduces a CME with reasonable velocity and has the potential to produce a CME event consistent with observation. Using this model in our future research, we will make some observation-based CME simulations. 

\begin{figure}[htpb!]
\begin{center}
    \includegraphics[width=0.5\linewidth,trim=1 1 1 1, clip]{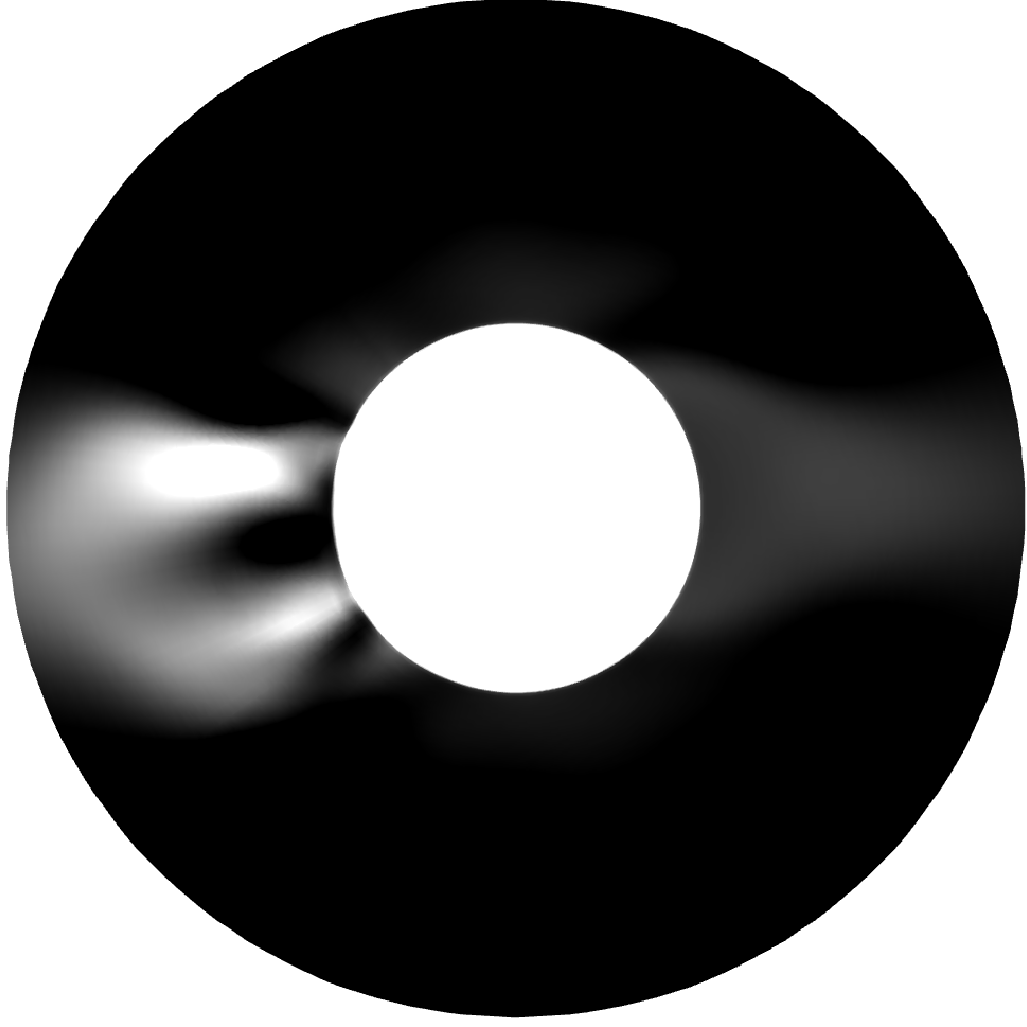}
\end{center}
  \caption{White-light pB image ranges from 1.4 to 4 $R_s$ on the meridian plane of $\phi_{\rm long}=250^{\circ}-70^{\circ}$, synthesised from the simulation result at 1 hour of CME simulation.}\label{CMEPb1hr}
\end{figure}

Additionally, we present the white-light pB image synthesised from the simulation result at 1 hour of the CME propagation process in Fig.~\ref{CMEPb1hr}. It is displayed on the same meridian plane as in Fig.~\ref{CMEsVratMeridians}. At this moment, the CME can be seen reaching 4 $R_s$ on the eastern limb of this meridian plane. It shows that the outline of the simulated CME shape is visible in the pB image, suggesting that observed pB images could be used to validate and refine the observation-based CME simulations in our future research.

As for the explicit coronal model SIP-EFVM, it breaks down around the 30th time step due to the appearance of negative thermal pressure. This may be attributed to the fact that the time-dependent $\mathbf{B}_1$ is no longer small after inserting the flux rope into the background corona and the 
 discretisation error in $\mathbf{B}_1$ leads to the appearance of such nonphysical thermal pressure. Since this issue only arises in the explicit model and requires significant additional effort for further verification, we intend to address it more thoroughly in future research.

  \subsection{An ad hoc simulation with very low plasma $\beta$}\label{sec:adhocsimulation}
   In this subsection, we describe the manufactured test we carried out by utilising the SIP-IFVM model to mimic a very low-$\beta$ problem. In the test simulation, we multiplied the initial potential field of CR 2219 by a factor of 5 and then employed the SIP-IFVM model to achieve the quasi-steady state coronal structure. 

\begin{figure*}[htpb!]
\begin{center}
  \vspace*{0.01\linewidth}
    \includegraphics[width=0.8\textwidth,trim=1 1 1 1, clip]{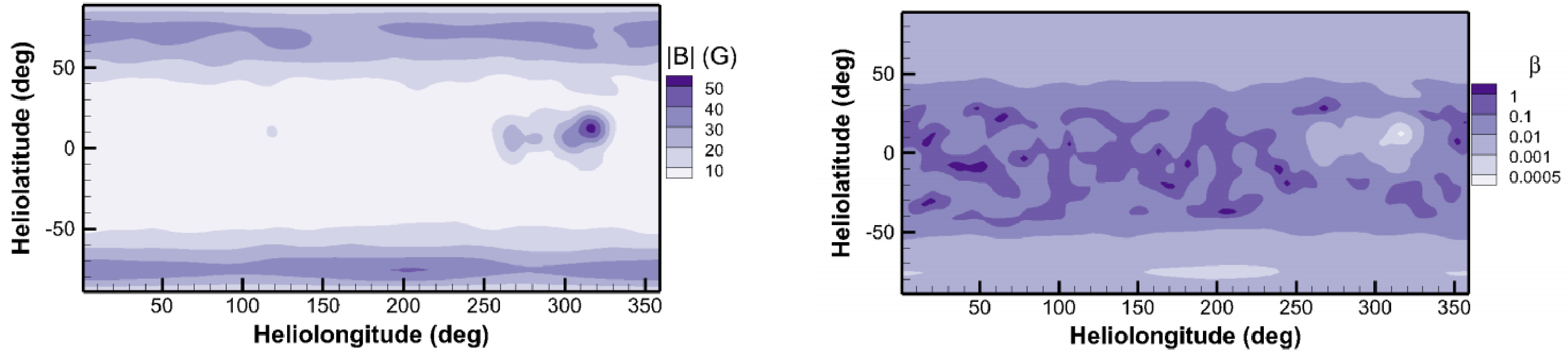}
\end{center}
\caption{Synoptic maps of the magnetic field strength in a unit of Gauss (left) and the plasma $\beta$ distribution (right) at $1.015 R_s$ for the test case with an enhanced magnetic field.}\label{PlasmabetaandMagstrength}
\end{figure*}

Figure~\ref{PlasmabetaandMagstrength} displays the synoptic map of the magnetic field (left) and the corresponding plasma $\beta$ (right) at a quasi-steady state near the solar surface. It can be seen that after enlarging the magnetic field strength, the local $\beta$ value can be as small as  $5 \times 10^{-4}$, and the magnetic field strength ranges from 5 to 50 Gauss in most regions near the solar surface. Moreover, it takes only $0.18$ hours to converge to the steady state. In the CME simulation, we enhanced the magnetic field of the flux roped with a factor of 2.5 and set $\chi=0.125$. It costs 0.67 hours to finish the time-dependent CME simulation of 6 hours of physical time. Compared with the simulation by small time steps (CFL=1), which cost 18.03 hours, this SIP-IFVM model is very efficient.

\begin{figure*}[htpb!]
\begin{center}\includegraphics[width=0.9\linewidth,trim=1 1 1 1, clip]{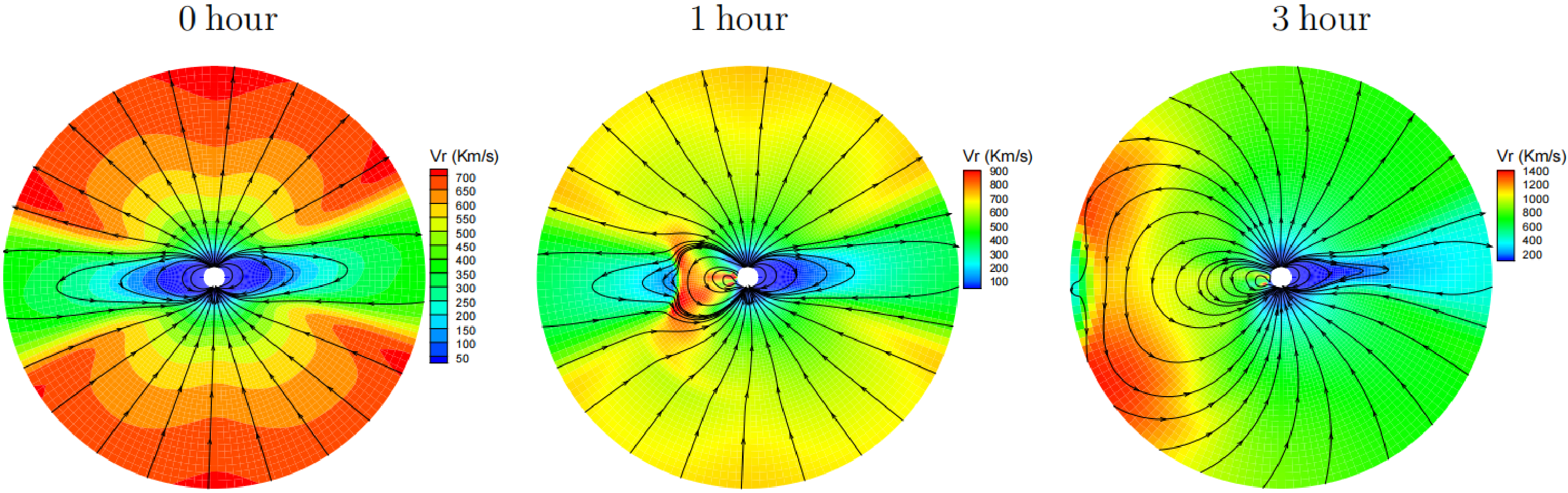}
\end{center}
  \caption{Magnetic field lines from 1 to $20 R_s$ overlaid on contours of the radial velocity on the meridian planes of $\phi_{\rm long}=250^{\circ}-70^{\circ}$ for the manufactured test with enhanced magnetic fields. The left, middle, and right panels correspond to the simulation results at 0, 1, and 3 hours of the CME simulation, respectively.}\label{CMEsVratMeridiansadhoc}
\end{figure*}

   In Fig.~\ref{CMEsVratMeridiansadhoc}, we further present snapshots of the radial speed $V_r$ and magnetic field lines at 0 (left), 1 (middle), and 3 (right) hours of the CME propagation process for the manufactured test with enhanced magnetic fields. These 2D modelled contours of radial velocity are superimposed with magnetic field lines and illustrated on the same meridian of Fig.~ \ref{CMEsVratMeridians}. It can be seen that a shock appears in this simulation, the volume overlaid with the flux rope magnetic field expands gradually. Though the topology of magnetic field lines changes more gradually, the radial velocity of this shock is faster. 

\section{Conclusions}\label{sec:conclusion}
   In this paper, we present our design of an MHD model of the solar corona and CME, with an efficient and time-accurate implicit strategy, called the Solar Interplanetary Phenomena-Implicit Finite Volume Method (SIP-IFVM) coronal model. 
 After extensive validation and evaluation, we  conclude that the SIP-IFVM coronal model has the following merits, providing strong justification for using a fully implicit scheme in time-dependent coronal and CME simulations.
    \begin{enumerate}
      \item  The SIP-IFVM coronal model is both time-accurate and highly computationally efficient. Adopting large time steps can still yield consistent results, comparable to those calculated at small time steps. Compared to the simulation using a small time-step size determined by the CFL condition, by adopting an appropriate large time-step size, the SIP-IFVM model achieves a six-fold speeding-up time in CME simulations covering 6 hours of physical time, with the average relative difference in plasma density, $\rm RD_{ave,\rho}$, being no more than $2.0\%$. In addition, by adopting a large time-step size, the implicit quasi-steady-state coronal model achieves more than a six-fold speed-up, with $\rm RD_{ave,\rho}$ being only $3.05\%$, compared to the explicit model. 
      The total wall-clock time of the quasi-steady coronal and time-dependent CME simulations is less than 0.6 hours (192 CPU cores, $\sim$ 1 M cells).

      \item  The SIP-IFVM coronal model can robustly and efficiently deal with time-dependent problems with extremely low plasma $\beta$ regions.
Compared to the simulation using a small time-step size determined by the CFL condition, by adopting an appropriate large time-step size, the SIP-IFVM model achieves a speedup of over 7 $\times$ in  ad hoc simulations, where the plasma $\beta$ can be as low as about $5 \times 10^{-4}$, with $\rm RD_{ave,\rho}$ being no more than $2.4\%$.
  
      \item  The SIP-IFVM coronal model can reproduce a quasi-steady state coronal structure consistent with observations, as well as simulate an explosive CME event with a reasonable evolution pattern of magnetic and flow fields. The relatively realistic simulation results have been achieved by adopting the thermodynamic MHD equations, which include the heat conduction term to account for energy exchanges. Additionally, the high flexibility in practical CME simulations is guaranteed by using the RBSL flux rope, which can trigger a CME event by introducing only the flux-rope magnetic field into the background corona. This allows the electric current path to take on an arbitrary shape.
   \end{enumerate}
   
   These simulation results demonstrate that the SIP-IFVM model is very efficient and numerically stable. It is a promising approach to simulating time-varying events in solar corona with low plasma $\beta$ in practical applications, in a  timely and accurate manner. In addition, we have also made some preliminary attempts to use the SIP-IFVM coronal model and observation-based RBSL flux rope to mimic a realistic CME event. The CME simulation results are consistent with the white-light pB images observed from COR1/STEREO-A/B and COR2/STEREO-A/B. We will continue conducting observation-based CME simulations using this model and will present our findings in future papers.

   Although this established solar coronal shows various merits and acts as a promising tool for reproducing the large-scale structures of the solar corona and timely and accurately simulating time-varying CME events in the solar corona, there is still room for further improvement. A proper modification of the Jacobian matrix in Eq. (\ref{uniformmodifiedglobalPseIBElinearized}), which reduces the mismatch between the residual and Jacobian matrix, may lead to a better convergence rate \cite[e.g.][]{Otero2015,XIA2014406}. Further research on the calibration of the coronal and CME model based on remote sensing and in situ observation is still worthwhile to make the SIP-IFVM model perform better in reproducing more realistic results. Extending the SIP-IFVM coronal model to a high-order accurate model may make it capable of performing high-fidelity simulations to capture subtle structures during the time-dependent coronal simulations. In addition, it may be worthwhile to start the global coronal simulation from some more consistent low coronal simulation results. For example, we could try to use the initial magnetic field above an active region calculated by the low coronal magnetic-friction (MF) model and the evolving electric field at the photosphere derived from a time series of observed photosphere magnetic field to drive detailed MHD simulations of active regions \citep{Hoeksema2020} in the SIP-IFVM global coronal model.
There are also some issues that are worth further discussion, and we will attempt to address these in our future research to further improve this model.
 \begin{enumerate}
     \item The pseudo-time iteration during each physical time step can reduce the computation efficiency. However, an appropriate physical time step size can help maintain required temporal accuracy without much reduction in computational efficiency; also, a proper pseudo-time step can also help accelerate the steady-state simulation's convergence rate in pseudo-time, $\tau$. Although the physical-time steps and pseudo-time steps used in this paper demonstrate a good performance, more effective and flexible time-step adaptation strategies may be possible. We will try to find a better plan for selecting time-step sizes in our future research works.   
     
     \item Considering that this paper is mainly aimed to extending the quasi-steady state coronal model to a time-dependent coronal model capable of efficiently simulating CMEs, we did not modify the governing equations in \cite{Wang_2022} and we retained $\gamma=1.05$, which is close to an isothermal process ($\gamma=1$). In the future, we will try to recover a value of $\gamma=\frac{5}{3}$ for the adiabatic process in coronal simulations after considering more thermodynamic mechanisms, such as radiative losses and more physically consistent heating source terms.
 \end{enumerate}

\begin{acknowledgements}
The authors thank Prof. Xueshang Feng, Dr. Xiaojing Liu, Dr. Man Zhang and Dr. Yuhao Zhou for their valuable comments. The work is jointly supported by the European Union’s Horizon 2020 research and innovation programme under a grant agreement
$\rm N^{\circ}$ 870405 (EUHFORIA 2.0) and National Natural Science Foundation of China (grant Nos. 42030204, 42074208 and 42104168).
This work has been granted by the  AFOSR basic research initiative project FA9550-18-1-0093.
This work is also part of a project supported by the Specialized Research Fund for State Key Laboratories, which is managed by the Chinese State Key Laboratory of Space Weather.
These results are also obtained in the framework of the projects C14/19/089 (C1 project Internal Funds KU Leuven), G0B5823N and G002523N(FWO-Vlaanderen), 4000134474 (SIDC Data Exploitation, ESA Prodex-12), and Belspo project B2/191/P1/SWiM. F.Z.\ is supported by the Research Council of Norway through its Centres of Excellence scheme, project number
262622. The resources and
services used in this work were provided by the VSC (Flemish Supercomputer Centre), funded by the Research Foundation – Flanders (FWO) and the Flemish Government.
This work utilises data obtained by the Global Oscillation Network Group (GONG) program, managed by the National Solar Observatory and operated by AURA, Inc., under a cooperative agreement with the National Science Foundation. The data were acquired by instruments operated by the Big Bear Solar Observatory, High Altitude Observatory, Learmonth Solar Observatory, Udaipur Solar Observatory, Instituto de Astrof{\'i}sica de Canarias, and Cerro Tololo Inter-American Observatory. This work utilises LASCO C2/SOHO and SDO/AIA data. The authors also acknowledge the use of the STEREO/SECCHI data produced by a consortium of the NRL (US), LMSAL (US), NASA/GSFC (US), RAL (UK), UBHAM (UK), MPS (Germany), CSL (Belgium), IOTA (France), and IAS (France). 
\end{acknowledgements}

\bibliographystyle{aa}
\bibliography{SIP_IFVM}

\begin{thebibliography}{169}
\expandafter\ifx\csname natexlab\endcsname\relax\def\natexlab#1{#1}\fi

\bibitem[{{Abbo} {et~al.}(2015){Abbo}, {Lionello}, {Riley}, \&
  {Wang}}]{Abbo2015}
{Abbo}, L., {Lionello}, R., {Riley}, P., \& {Wang}, Y.~M. 2015, Sol. Phys.,
  290, 2043

\bibitem[{{Abramenko} {et~al.}(2010){Abramenko}, {Yurchyshyn}, {Linker},
  {Miki{\'c}}, {Luhmann}, \& {Lee}}]{Abramenko2010}
{Abramenko}, V., {Yurchyshyn}, V., {Linker}, J.~A., {et~al.} 2010, ApJ, 712,
  813

\bibitem[{{Arge} {et~al.}(2004){Arge}, {Luhmann}, {Odstrcil}, {Schrijver}, \&
  {Li}}]{ARGE20041295}
{Arge}, C., {Luhmann}, J., {Odstrcil}, D., {Schrijver}, C., \& {Li}, Y. 2004,
  J. Atmos. Sol.-Terr. Phys., 66, 1295

\bibitem[{{Arge} {et~al.}(2003){Arge}, {Odstrcil}, {Pizzo}, \&
  {Mayer}}]{Arge2003ImprovedMF}
{Arge}, C.~N., {Odstrcil}, D., {Pizzo}, V.~J., \& {Mayer}, L.~R. 2003, AIP
  Conf. Proc., 679, 190

\bibitem[{{Baker}(1998)}]{BAKER19987}
{Baker}, D.~N. 1998, Adv. Space Res, 22, 7

\bibitem[{{Balsara}(2010)}]{Balsara20101970}
{Balsara}, D.~S. 2010, J. Comput. Phys., 229, 1970

\bibitem[{{Barth}(1991)}]{BARTH1991}
{Barth}, T.~J. 1991, in 10th Computational Fluid Dynamics Conference, 24 June
  1991-26 June 1991, Honolulu, HI, U.S.A., aIAA 1991-1548

\bibitem[{{Barth}(1993)}]{BARTH1993}
{Barth}, T.~J. 1993, in 31st Aerospace Sciences Meeting, aIAA 1993-668

\bibitem[{{Barth} \& {Jespersen}(1989)}]{Barth1989}
{Barth}, T.~J. \& {Jespersen}, D.~C. 1989, in 27th Aerospace Sciences Meeting,
  aIAA 1989-0366

\bibitem[{{Bijl} {et~al.}(2002){Bijl}, {Carpenter}, {Vatsa}, \&
  {Kennedy}}]{BIJL2002313}
{Bijl}, H., {Carpenter}, M.~H., {Vatsa}, V.~N., \& {Kennedy}, C.~A. 2002, J.
  Comput. Phys., 179, 313

\bibitem[{Bourdin(2017)}]{Bourdin2017}
Bourdin, P.-A. 2017, ApJL, 850, L29

\bibitem[{{Brchnelova} {et~al.}(2022){Brchnelova}, {Ku{\'{z}}ma}, {Perri},
  {Lani}, \& {Poedts}}]{Brchnelova_2022}
{Brchnelova}, M., {Ku{\'{z}}ma}, B., {Perri}, B., {Lani}, A., \& {Poedts}, S.
  2022, ApJS, 263, 18

\bibitem[{{Brchnelova} {et~al.}(2023){Brchnelova}, {Ku{\'{z}}ma}, {Zhang},
  {Lani}, \& {Poedts}}]{brchnelova2023role}
{Brchnelova}, M., {Ku{\'{z}}ma}, B., {Zhang}, F., {Lani}, A., \& {Poedts}, S.
  2023, A \& A, 676

\bibitem[{{Brueckner} {et~al.}(1995){Brueckner}, {Howard}, {Koomen},
  {Korendyke}, {Michels}, {Moses}, {Socker}, {Dere}, {Lamy}, {Llebaria},
  {Bout}, {Schwenn}, {Simnett}, {Bedford}, \& {Eyles}}]{Brueckner1995}
{Brueckner}, G.~E., {Howard}, R.~A., {Koomen}, M.~J., {et~al.} 1995, Sol.
  Phys., 162, 357

\bibitem[{{Burlaga} {et~al.}(1981){Burlaga}, {Sittler}, {Mariani}, \&
  {Schwenn}}]{Burlaga1981}
{Burlaga}, L.~F., {Sittler}, E.~C., {Mariani}, F., \& {Schwenn}, R. 1981, J.
  Geophys. Res.: Space Phys., 86, 6673

\bibitem[{Caplan {et~al.}(2019)Caplan, Linker, Miki{\'c}, Downs, T{\"o}r{\"o}k,
  \& Titov}]{Caplan_2019}
Caplan, R.~M., Linker, J.~A., Miki{\'c}, Z., {et~al.} 2019, J. Phys.: Conf.
  Ser., 1225, 012012

\bibitem[{Caplan {et~al.}(2017)Caplan, Miki{\'c}, Linker, \&
  Lionello}]{Caplan_2017}
Caplan, R.~M., Miki{\'c}, Z., Linker, J.~A., \& Lionello, R. 2017, J. Phys.:
  Conf. Ser., 837, 012016

\bibitem[{{Chen}(2011)}]{Chen2011}
{Chen}, P.~F. 2011, Living Rev. Sol. Phys., 8, 1

\bibitem[{Cheng {et~al.}(2017)Cheng, Guo, \& Ding}]{Chen2017}
Cheng, X., Guo, Y., \& Ding, M. 2017, Sci. China Earth Sci., 60

\bibitem[{{Cheung} \& {DeRosa}(2012)}]{Cheung2012}
{Cheung}, M. C.~M. \& {DeRosa}, M.~L. 2012, ApJ, 757, 147

\bibitem[{{Detman} {et~al.}(2006){Detman}, {Smith}, {Dryer}, {Fry}, {Arge}, \&
  {Pizzo}}]{Detman2005}
{Detman}, T., {Smith}, Z., {Dryer}, M., {et~al.} 2006, J. Geophys. Res.: Space
  Phys., 111, A07102

\bibitem[{{Dryer}(2007)}]{Dryer2007nodoc}
{Dryer}, M. 2007, Asian J. Phys., 16, 97

\bibitem[{{Einfeldt} {et~al.}(1991){Einfeldt}, {Munz}, {Roe}, \&
  {Sjogreen}}]{EINFELDT1991273}
{Einfeldt}, B., {Munz}, C.~D., {Roe}, P.~L., \& {Sjogreen}, B. 1991, J. Comput.
  Phys., 92, 273

\bibitem[{{Endeve} {et~al.}(2003){Endeve}, {Leer}, \& {Holzer}}]{Endeve_2003}
{Endeve}, E., {Leer}, E., \& {Holzer}, T.~E. 2003, ApJ, 589, 1040

\bibitem[{{Feng}(2020{\natexlab{a}})}]{Feng2020chapt2}
{Feng}, X.~S. 2020{\natexlab{a}}, in Magnetohydrodynamic Modeling of the Solar
  Corona and Heliosphere, 125--337

\bibitem[{{Feng}(2020{\natexlab{b}})}]{Feng2020book}
{Feng}, X.~S. 2020{\natexlab{b}}, Magnetohydrodynamic Modeling of the Solar
  Corona and Heliosphere (Singapore: Springer)

\bibitem[{{Feng} {et~al.}(2012{\natexlab{a}}){Feng}, {Jiang}, {Xiang}, {Zhao},
  \& {Wu}}]{Feng_2012}
{Feng}, X.~S., {Jiang}, C.~W., {Xiang}, C.~Q., {Zhao}, X.~P., \& {Wu}, S.~T.
  2012{\natexlab{a}}, ApJ, 758, 62

\bibitem[{{Feng} {et~al.}(2017){Feng}, {Li}, {Xiang}, {Zhang}, {Li}, \&
  {Wei}}]{Feng_2017}
{Feng}, X.~S., {Li}, C.~X., {Xiang}, C.~Q., {et~al.} 2017, ApJS, 233, 10

\bibitem[{{Feng} {et~al.}(2019){Feng}, {Liu}, {Xiang}, {Li}, \&
  {Wei}}]{FengandLiu2019}
{Feng}, X.~S., {Liu}, X.~J., {Xiang}, C.~Q., {Li}, H.~C., \& {Wei}, F.~S. 2019,
  ApJ, 871, 226

\bibitem[{{Feng} {et~al.}(2015){Feng}, {Ma}, \& {Xiang}}]{FengMa2015}
{Feng}, X.~S., {Ma}, X.~P., \& {Xiang}, C.~Q. 2015, J. Geophys. Res.: Space
  Phys., 120, 10,159

\bibitem[{{Feng} {et~al.}(2021){Feng}, {Wang}, Xiang, Liu, {Zhang}, {Zhao}, \&
  {Shen}}]{Feng_2021}
{Feng}, X.~S., {Wang}, H.~P., Xiang, C.~Q., {et~al.} 2021, ApJS, 257, 34

\bibitem[{{Feng} {et~al.}(2011{\natexlab{a}}){Feng}, {Xiang}, \&
  {Zhong}}]{Feng_2011Chinese}
{Feng}, X.~S., {Xiang}, C.~Q., \& {Zhong}, D.~K. 2011{\natexlab{a}}, Sci
  Sin-Terrae, 41, 1

\bibitem[{{Feng} {et~al.}(2013){Feng}, {Xiang}, \& {Zhong}}]{Feng_2013Chinese}
{Feng}, X.~S., {Xiang}, C.~Q., \& {Zhong}, D.~K. 2013, Sci Sin-Terrae, 43, 912

\bibitem[{{Feng} {et~al.}(2014{\natexlab{a}}){Feng}, {Xiang}, {Zhong}, {Zhou},
  {Yang}, \& {Ma}}]{FENG20141965}
{Feng}, X.~S., {Xiang}, C.~Q., {Zhong}, D.~K., {et~al.} 2014{\natexlab{a}},
  Comput. Phys. Commun, 185, 1965

\bibitem[{{Feng} {et~al.}(2012{\natexlab{b}}){Feng}, {Yang}, {Xiang}, {Jiang},
  {Ma}, {Wu}, {Zhong}, \& {Zhou}}]{Feng2012}
{Feng}, X.~S., {Yang}, L.~P., {Xiang}, C.~Q., {et~al.} 2012{\natexlab{b}}, Sol.
  Phys, 279, 207

\bibitem[{{Feng} {et~al.}(2010){Feng}, {Yang}, {Xiang}, {Wu}, {Zhou}, \&
  {Zhong}}]{Feng_2010}
{Feng}, X.~S., {Yang}, L.~P., {Xiang}, C.~Q., {et~al.} 2010, ApJ, 723, 300

\bibitem[{{Feng} {et~al.}(2014{\natexlab{b}}){Feng}, {Zhang}, \&
  {Zhou}}]{Feng_2014}
{Feng}, X.~S., {Zhang}, M., \& {Zhou}, Y.~F. 2014{\natexlab{b}}, ApJS, 214, 6

\bibitem[{{Feng} {et~al.}(2011{\natexlab{b}}){Feng}, {Zhang}, {Xiang}, {Yang},
  {Jiang}, \& {Wu}}]{Feng_2011}
{Feng}, X.~S., {Zhang}, S.~H., {Xiang}, C.~Q., {et~al.} 2011{\natexlab{b}},
  ApJ, 734, 50

\bibitem[{{Feng} {et~al.}(2007){Feng}, {Zhou}, \& {Wu}}]{Feng_2007}
{Feng}, X.~S., {Zhou}, Y.~F., \& {Wu}, S.~T. 2007, ApJ, 655, 1110

\bibitem[{Fr{\"a}nz \& Harper(2002)}]{FRANZ2002217}
Fr{\"a}nz, M. \& Harper, D. 2002, Planet. Space Sci, 50, 217

\bibitem[{{Frazin} {et~al.}(2007){Frazin}, {V{\'{a}}squez}, {Kamalabadi}, \&
  {Park}}]{Frazin2007}
{Frazin}, R.~A., {V{\'{a}}squez}, A.~M., {Kamalabadi}, F., \& {Park}, H. 2007,
  ApJ, 671, 201

\bibitem[{{Fuchs} {et~al.}(2010){Fuchs}, {McMurry}, {Mishra}, {Risebro}, \&
  {Waagan}}]{FUCHS2010JCP}
{Fuchs}, F.~G., {McMurry}, A.~D., {Mishra}, S., {Risebro}, N.~H., \& {Waagan},
  K. 2010, J. Comput. Phys., 229, 4033

\bibitem[{{Fuchs} {et~al.}(2009){Fuchs}, {Mishra}, \& {Risebro}}]{Fuchs2009}
{Fuchs}, F.~G., {Mishra}, S., \& {Risebro}, N.~H. 2009, J. Comput. Phys., 228,
  641

\bibitem[{{Gibson} \& {Low}(1998)}]{Gibson_1998}
{Gibson}, S.~E. \& {Low}, B.~C. 1998, ApJ, 493, 460

\bibitem[{Godunov(1959)}]{Godunov1959Adifference}
Godunov, S.~K. 1959, Mat. Sb. (N.S.), 1959,, 271

\bibitem[{Gombosi {et~al.}(2018)Gombosi, {Van der Holst}, Manchester, \&
  Sokolov}]{Gombosi2018}
Gombosi, T.~I., {Van der Holst}, B., Manchester, W.~B., \& Sokolov, I.~V. 2018,
  Living Rev. Sol. Phys., 15, 4

\bibitem[{Goodrich {et~al.}(2004)Goodrich, Sussman, Lyon, Shay, \&
  Cassak}]{GOODRICH20041469}
Goodrich, C., Sussman, A., Lyon, J., Shay, M., \& Cassak, P. 2004, J. Atmos.
  Sol.-Terr. Phys., 66, 1469, towards an Integrated Model of the Space Weather
  System

\bibitem[{{Groth} {et~al.}(2000){Groth}, {De Zeeuw}, {Gombosi}, \&
  {Powell}}]{Groth2000}
{Groth}, C.~P.~T., {De Zeeuw}, D.~L., {Gombosi}, T.~I., \& {Powell}, K.~G.
  2000, J. Geophys. Res.: Space Phys., 105, 25053

\bibitem[{Guo {et~al.}(2023)Guo, Linan, Poedts, Guo, Lani, Schmieder,
  Brchnelova, Perri, Baratashvili, Ni, \& Chen}]{guo2023}
Guo, J.~H., Linan, L., Poedts, S., {et~al.} 2023, A \& A

\bibitem[{Guo {et~al.}(2021)Guo, Ni, Qiu, Zhong, Guo, \& Chen}]{GUO202108}
Guo, J.~H., Ni, Y.~W., Qiu, Y., {et~al.} 2021, ApJ, 917, 81

\bibitem[{{Guo} {et~al.}(2023){Guo}, {Ni}, {Zhong}, {Guo}, {Xia}, {Li},
  {Poedts}, {Schmieder}, \& {Chen}}]{Guo2023b}
{Guo}, J.~H., {Ni}, Y.~W., {Zhong}, Z., {et~al.} 2023, ApJS, 266, 3

\bibitem[{{Guo}(2015)}]{Guo2015}
{Guo}, X.~C. 2015, J. Comput. Phys., 290, 352

\bibitem[{Guo {et~al.}(2017)Guo, Cheng, \& Ding}]{Guo2017}
Guo, Y., Cheng, X., \& Ding, M. 2017, Sci. China Earth Sci., 60

\bibitem[{Guo {et~al.}(2016)Guo, Xia, Keppens, \& Valori}]{Guo_2016}
Guo, Y., Xia, C., Keppens, R., \& Valori, G. 2016, ApJ, 828, 82

\bibitem[{{Guo} {et~al.}(2019){Guo}, {Xu}, {Ding}, {Chen}, {Xia}, \&
  {Keppens}}]{Guo2019}
{Guo}, Y., {Xu}, Y., {Ding}, M.~D., {et~al.} 2019, ApJL, 884, L1

\bibitem[{Hamada {et~al.}(2018)Hamada, Asikainen, Virtanen, \&
  Mursula}]{Hamada2018}
Hamada, A., Asikainen, T., Virtanen, I., \& Mursula, K. 2018, Sol. Phys., 293,
  71

\bibitem[{{Hayashi} {et~al.}(2021){Hayashi}, {Abbett}, {Cheung}, \&
  {Fisher}}]{Hayashi_2021}
{Hayashi}, K., {Abbett}, W.~P., {Cheung}, M. C.~M., \& {Fisher}, G.~H. 2021,
  ApJS, 254, 1

\bibitem[{{Hayashi} {et~al.}(2006{\natexlab{a}}){Hayashi}, {Benevolenskaya},
  {Hoeksema}, {Liu}, \& {Zhao}}]{Hayashi2006}
{Hayashi}, K., {Benevolenskaya}, E., {Hoeksema}, T., {Liu}, Y., \& {Zhao},
  X.~P. 2006{\natexlab{a}}, ApJ, 636, L165

\bibitem[{{Hayashi} {et~al.}(2006{\natexlab{b}}){Hayashi}, {Zhao}, \&
  {Liu}}]{Hayashi2006CONEFR}
{Hayashi}, K., {Zhao}, X.~P., \& {Liu}, Y. 2006{\natexlab{b}}, Geophys. Res.
  Lett., 33

\bibitem[{{Hayes} {et~al.}(2001){Hayes}, {Vourlidas}, \& {Howard}}]{Hayes2001}
{Hayes}, A.~P., {Vourlidas}, A., \& {Howard}, R.~A. 2001, ApJ, 548, 1081

\bibitem[{{Hoeksema} {et~al.}(2020){Hoeksema}, {Abbett}, {Bercik}, {Cheung},
  {DeRosa}, {Fisher}, {Hayashi}, {Kazachenko}, {Liu}, {Lumme}, {Lynch}, {Sun},
  \& {Welsch}}]{Hoeksema2020}
{Hoeksema}, J.~T., {Abbett}, W.~P., {Bercik}, D.~J., {et~al.} 2020, ApJS, 250,
  28

\bibitem[{{Hollweg}(1978)}]{Hollweg1978}
{Hollweg}, J.~V. 1978, Rev. Geophys., 16, 689

\bibitem[{{Hoshyari} {et~al.}(2020){Hoshyari}, {Mirzaee}, \&
  {Ollivier-Gooch}}]{Hoshyari_2020}
{Hoshyari}, S., {Mirzaee}, E., \& {Ollivier-Gooch}, C. 2020, AIAA J., 58, 1490

\bibitem[{Howard {et~al.}(2008)Howard, Moses, Vourlidas, Newmark, Socker,
  Plunkett, Korendyke, Cook, Hurley, Davila, Thompson, St~Cyr, Mentzell,
  Mehalick, Lemen, Wuelser, Duncan, Tarbell, Wolfson, Moore, Harrison, Waltham,
  Lang, Davis, Eyles, Mapson-Menard, Simnett, Halain, Defise, Mazy, Rochus,
  Mercier, Ravet, Delmotte, Auchere, Delaboudiniere, Bothmer, Deutsch, Wang,
  Rich, Cooper, Stephens, Maahs, Baugh, McMullin, \& Carter}]{Howard2008}
Howard, R.~A., Moses, J.~D., Vourlidas, A., {et~al.} 2008, Space Sci. Rev.,
  136, 67

\bibitem[{Jiang {et~al.}(2010)Jiang, Feng, Zhang, \& Zhong}]{Jiang2010}
Jiang, C.~W., Feng, X.~S., Zhang, J., \& Zhong, D.~K. 2010, Sol. Phys., 267,
  463

\bibitem[{Jiang {et~al.}(2016)Jiang, Wu, Feng, \& Hu}]{Jiang201605}
Jiang, C.~W., Wu, S.~T., Feng, X.~S., \& Hu, Q. 2016, Nat. Commun., 7

\bibitem[{{Jiang} \& {Wu}(1999)}]{Jiang1999}
{Jiang}, G.-S. \& {Wu}, C.-C. 1999, J. Comput. Phys., 150, 561

\bibitem[{Jin {et~al.}(2017)Jin, Manchester, van~der Holst, Sokolov, Toth,
  Mullinix, Taktakishvili, Chulaki, \& Gombosi}]{Jin_2017}
Jin, M., Manchester, W.~B., van~der Holst, B., {et~al.} 2017, ApJ, 834, 173

\bibitem[{{Kaiser} {et~al.}(2008){Kaiser}, {Kucera}, {Davila}, {Cyr},
  {Guhathakurta}, \& {Christian}}]{Kaiser2008TheSM}
{Kaiser}, M.~L., {Kucera}, T.~A., {Davila}, J.~M., {et~al.} 2008, Space Sci.
  Rev., 136, 5

\bibitem[{Kataoka {et~al.}(2009)Kataoka, Ebisuzaki, Kusano, Shiota, Inoue,
  Yamamoto, \& Tokumaru}]{Kataoka2009}
Kataoka, R., Ebisuzaki, T., Kusano, K., {et~al.} 2009, J. Geophys. Res.: Space
  Phys., 114

\bibitem[{Keppens {et~al.}(2023)Keppens, Braileanu, Zhou, Ruan, Xia, Guo,
  Claes, \& Bacchini}]{Keppens2023}
Keppens, R., Braileanu, B.~P., Zhou, Y.~H., {et~al.} 2023, A \& A, 673

\bibitem[{{Koskinen} {et~al.}(2017){Koskinen}, {Baker}, {Balogh}, {Gombosi},
  {Veronig}, \& {Rudolf}}]{Koskinen2017}
{Koskinen}, H.~E.~J., {Baker}, D.~N., {Balogh}, A., {et~al.} 2017, Space Sci.
  Rev, 212, 1137

\bibitem[{{Ku{\'{z}}ma} {et~al.}(2023){Ku{\'{z}}ma}, {Brchnelova}, {Perri},
  {Baratashvili}, {Zhang}, {Lani}, \& {Poedts}}]{Kuzma_2023}
{Ku{\'{z}}ma}, B., {Brchnelova}, M., {Perri}, B., {et~al.} 2023, ApJ, 942, 31

\bibitem[{{Lemen} {et~al.}(2012){Lemen}, {Title}, {Akin}, {Boerner}, {Chou},
  {Drake}, {Duncan}, {Edwards}, {Friedlaender}, {Heyman}, {Hurlburt}, {Katz},
  {Kushner}, {Levay}, {Lindgren}, {Mathur}, {McFeaters}, {Mitchell}, {Rehse},
  {Schrijver}, {Springer}, {Stern}, {Tarbell}, {Wuelser}, {Wolfson}, {Yanari},
  {Bookbinder}, {Cheimets}, {Caldwell}, {Deluca}, {Gates}, {Golub}, {Park},
  {Podgorski}, {Bush}, {Scherrer}, {Gummin}, {Smith}, {Auker}, {Jerram},
  {Pool}, {Soufli}, {Windt}, {Beardsley}, {Clapp}, {Lang}, \&
  {Waltham}}]{Lemen2012}
{Lemen}, J.~R., {Title}, A.~M., {Akin}, D.~J., {et~al.} 2012, Sol. Phys., 275,
  17

\bibitem[{Li {et~al.}(2005)Li, Habbal, Li, \& Mountford}]{Li2005}
Li, B., Habbal, S.~R., Li, X., \& Mountford, C. 2005, J. Geophys. Res.: Space
  Phys., 110

\bibitem[{{Li} \& {Feng}(2018)}]{Lihuichao2018}
{Li}, H.~C. \& {Feng}, X.~S. 2018, J. Geophys. Res.: Space Phys., 123, 4488

\bibitem[{{Li} {et~al.}(2020){Li}, {Feng}, \& {Wei}}]{LiHuichao2020}
{Li}, H.~C., {Feng}, X.~S., \& {Wei}, F.~S. 2020, J. Space Weather Space Clim.

\bibitem[{{Li} {et~al.}(2021){Li}, {Feng}, \& {Wei}}]{LiHuichao2021}
{Li}, H.~C., {Feng}, X.~S., \& {Wei}, F.~S. 2021, J. Geophys. Res.: Space
  Phys., 126, e2020JA028870

\bibitem[{{Li} {et~al.}(2019){Li}, {Lou}, {Luo}, \& {Nishikawa}}]{Lingquan2019}
{Li}, L.~Q., {Lou}, J.~L., {Luo}, H., \& {Nishikawa}, H. 2019, in AIAA Aviation
  2019 Forum, aIAA 2019-3060

\bibitem[{Li \& Ren(2012)}]{Li2012}
Li, W.~A. \& Ren, Y.-X. 2012, J. Comput. Phys., 231, 4053¡V4077

\bibitem[{Li {et~al.}(2011)Li, Ren, Lei, \& Luo}]{Li2011}
Li, W.~A., Ren, Y.-X., Lei, G.~D., \& Luo, H. 2011, J. Comput. Phys., 230,
  7775¡V7795

\bibitem[{{Li} \& {Ren}(2012)}]{Li2012ijnm}
{Li}, W.~N. \& {Ren}, Y.~X. 2012, Int. J. Numer. Methods Fluids, 70, 742

\bibitem[{Linan {et~al.}(2023)Linan, Regnault, Perri, Brchnelova,
  {Ku{\'{z}}ma}, Lani, Poedts, \& Schmieder}]{Linan_2023}
Linan, L., Regnault, F., Perri, B., {et~al.} 2023, A \& A, 675, A101

\bibitem[{{Linker} {et~al.}(1999){Linker}, {Miki{\'c}}, {Biesecker}, {Forsyth},
  {Gibson}, {Lazarus}, {Lecinski}, {Riley}, {Szabo}, \&
  {Thompson}}]{Linker1999JGR}
{Linker}, J.~A., {Miki{\'c}}, Z., {Biesecker}, D.~A., {et~al.} 1999, J.
  Geophys. Res.: Space Phys., 104, 9809

\bibitem[{{Linker} {et~al.}(2024){Linker}, {Torok}, {Downs}, {Caplan}, {Titov},
  {Reyes}, {Lionello}, \& {Riley}}]{Linker2024}
{Linker}, J.~A., {Torok}, T., {Downs}, C., {et~al.} 2024, J. Phys.: Conf. Ser.,
  2742, 012012

\bibitem[{{Lionello} {et~al.}(2008){Lionello}, {Linker}, \&
  {Miki{\'{c}}}}]{Lionello_2008}
{Lionello}, R., {Linker}, J.~A., \& {Miki{\'{c}}}, Z. 2008, ApJ, 690, 902

\bibitem[{{Liu} {et~al.}(2023){Liu}, {Feng}, {Zhang}, \& {Zhao}}]{Liu_2023}
{Liu}, X.~J., {Feng}, X.~S., {Zhang}, M., \& {Zhao}, J.~M. 2023, ApJS, 265, 19

\bibitem[{{Liu} {et~al.}(2016){Liu}, {Zhang}, {Jiang}, \& {Ye}}]{LIU20161096}
{Liu}, Y.~L., {Zhang}, W.~W., {Jiang}, Y.~W., \& {Ye}, Z.~Y. 2016, Comput.
  Math. with Appl., 72, 1096

\bibitem[{{Ljubomir} \& {Larisa}(2012)}]{LjubomirNikolic2012}
{Ljubomir}, N. \& {Larisa}, T. 2012, IJGEE, 6, 698

\bibitem[{{Lugaz} \& {Roussev}(2011)}]{LUGAZ20111187}
{Lugaz}, N. \& {Roussev}, I. 2011, J. Atmos. Sol.-Terr. Phys., 73, 1187, three
  dimensional aspects of CMEs, their source regions and interplanetary
  manifestations

\bibitem[{{Luo} {et~al.}(2001){Luo}, {Baum}, \& {L{\"o}hner}}]{LUO2001137}
{Luo}, H., {Baum}, J.~D., \& {L{\"o}hner}, R. 2001, Comput. Fluids, 30, 137

\bibitem[{Marubashi {et~al.}(2016)Marubashi, Cho, Kim, Kim, Park, \&
  Ishibashi}]{Marubashi2016}
Marubashi, K., Cho, K., Kim, R.-S., {et~al.} 2016, Earth Planets Space, 68

\bibitem[{Mays {et~al.}(2015)Mays, Taktakishvili, Pulkkinen, Macneice,
  Rastaetter, Odstrcil, Jian, Richardson, Lasota, Zheng, \&
  Kuznetsova}]{Mays2015}
Mays, M., Taktakishvili, A., Pulkkinen, A., {et~al.} 2015, Sol. Phys., 290

\bibitem[{Mignone {et~al.}(2011)Mignone, Zanni, Tzeferacos, van Straalen,
  Colella, \& Bodo}]{Mignone_2012}
Mignone, A., Zanni, C., Tzeferacos, P., {et~al.} 2011, ApJS, 198, 7

\bibitem[{{Miki{\'c}} {et~al.}(1999){Miki{\'c}}, {Linker}, {Schnack},
  {Lionello}, \& {Tarditi}}]{Mikic1999}
{Miki{\'c}}, Z., {Linker}, J.~A., {Schnack}, D.~D., {Lionello}, R., \&
  {Tarditi}, A. 1999, Phys. Plasmas, 6, 2217

\bibitem[{{Nakamizo} {et~al.}(2009){Nakamizo}, {Tanaka}, {Kubo}, {Kamei},
  {Shimazu}, \& {Shinagawa}}]{Nakamizoetal2009}
{Nakamizo}, A., {Tanaka}, T., {Kubo}, Y., {et~al.} 2009, J. Geophys. Res.:
  Space Phys., 114

\bibitem[{{Newkirk} {et~al.}(1970){Newkirk}, {Dupree}, \&
  {Schmhl}}]{Newkirk1970}
{Newkirk}, G., {Dupree}, R.~G., \& {Schmhl}, E.~J. 1970, Sol. Phys., 15, 15

\bibitem[{Nieves-Chinchilla {et~al.}(2018)Nieves-Chinchilla, Linton, Hidalgo,
  \& Vourlidas}]{Nieves_Chinchilla_2018}
Nieves-Chinchilla, T., Linton, M.~G., Hidalgo, M.~A., \& Vourlidas, A. 2018,
  ApJ, 861, 139

\bibitem[{Noventa {et~al.}(2020)Noventa, Massa, Rebay, Bassi, \&
  Ghidoni}]{NOVENTA2020104529}
Noventa, G., Massa, F., Rebay, S., Bassi, F., \& Ghidoni, A. 2020, Comput.
  Fluids, 204, 104529

\bibitem[{Odstrcil \& Pizzo(1999)}]{Odstrcil1999}
Odstrcil, D. \& Pizzo, V.~J. 1999, J. Geophys. Res.: Space Phys., 104, 483

\bibitem[{{Odstrcil} {et~al.}(2004){Odstrcil}, {Pizzo}, {Linker}, {Riley},
  {Lionello}, \& {Miki{\'c}}}]{ODSTRCIL20041311}
{Odstrcil}, D., {Pizzo}, V.~J., {Linker}, J.~A., {et~al.} 2004, J. Atmos.
  Sol.-Terr. Phys., 66, 1311, towards an Integrated Model of the Space Weather
  System

\bibitem[{{Ogino} \& {Walker}(1984)}]{OginoTatsuki1984}
{Ogino}, T. \& {Walker}, R.~J. 1984, Geophys. Res. Lett., 11, 1018

\bibitem[{Orszag \& Tang(1979)}]{orszag_tang_1979}
Orszag, S.~A. \& Tang, C.-M. 1979, J. Fluid Mech., 90, 129

\bibitem[{{Otero} \& {Eliasson}(2015{\natexlab{a}})}]{OteroAcc2015}
{Otero}, E. \& {Eliasson}, P. 2015{\natexlab{a}}, Int J Comut Fluid Dyn, 29,
  133

\bibitem[{{Otero} \& {Eliasson}(2015{\natexlab{b}})}]{Otero2015}
{Otero}, E. \& {Eliasson}, P. 2015{\natexlab{b}}, Int J Comut Fluid Dyn, 29,
  313

\bibitem[{{Ouyang} {et~al.}(2017){Ouyang}, {Zhou}, {Chen}, \&
  {Fang}}]{Ouyang2017}
{Ouyang}, Y., {Zhou}, Y.~H., {Chen}, P.~F., \& {Fang}, C. 2017, ApJ, 835, 94

\bibitem[{Owens {et~al.}(2017)Owens, Lockwood, \& Riley}]{Owens2017}
Owens, M.~J., Lockwood, M., \& Riley, P. 2017, Sci Rep, 7, 41548

\bibitem[{{Parker}(1963)}]{Parker1963}
{Parker}, E.~N. 1963, {Interplanetary dynamical processes} (New York:
  Interscience Publishers)

\bibitem[{{Perri} {et~al.}(2023){Perri}, {Ku{\'{z}}ma}, {Brchnelova},
  {Baratashvili}, {Zhang}, {Leitner}, {Lani}, \& {Poedts}}]{Perri_2023}
{Perri}, B., {Ku{\'{z}}ma}, B., {Brchnelova}, M., {et~al.} 2023, ApJ, 943, 124

\bibitem[{{Perri} {et~al.}(2022){Perri}, {Leitner}, {Brchnelova},
  {Baratashvili}, {Ku{\'{z}}ma}, {Zhang}, {Lani}, \& {Poedts}}]{Perri_2022}
{Perri}, B., {Leitner}, P., {Brchnelova}, M., {et~al.} 2022, ApJ, 936, 19

\bibitem[{{Persson}(2013)}]{PERSSON2013414}
{Persson}, P.-O. 2013, J. Comput. Phys., 233, 414

\bibitem[{{Petrie} {et~al.}(2011){Petrie}, {Canou}, \&
  {Amari}}]{Petrie2011SoPh}
{Petrie}, G.~J.~D., {Canou}, A., \& {Amari}, T. 2011, Sol. Phys., 274, 163

\bibitem[{{Petrov} {et~al.}(2017){Petrov}, {Titarev}, {Utyuzhnikov}, \&
  {Chikitkin}}]{Petrov2017}
{Petrov}, M.~N., {Titarev}, V.~A., {Utyuzhnikov}, S.~V., \& {Chikitkin}, A.~V.
  2017, Comput. Math. Math. Phys., 57, 1856

\bibitem[{{Poedts, S.} {et~al.}(2020){Poedts, S.}, {Lani, A.}, {Scolini, C.},
  {Verbeke, C.}, {Wijsen, N.}, {Lapenta, G.}, {Laperre, B.}, {Millas, D.},
  {Innocenti, M. E.}, {Chane, E.}, {Baratashvili, T.}, {Samara, E.}, {Van der
  Linden, R.}, {Rodriguez, L.}, {Vanlommel, P.}, {Vainio, R.}, {Afanasiev, A.},
  {Kilpua, E.}, {Pomoell, J.}, {Sarkar, R.}, {Aran, A.}, {Sanahuja, B.},
  {Paredes, J. M.}, {Clarke, E.}, {Thomson, A.}, {Rouilard, A.}, {Pinto, R.
  F.}, {Marchaudon, A.}, {Blelly, P.-L.}, {Gorce, B.}, {Plotnikov, I.},
  {Kouloumvakos, A.}, {Heber, B.}, {Herbst, K.}, {Kochanov, A.}, {Raeder, J.},
  \& {Depauw, J.}}]{Poedts_2020}
{Poedts, S.}, {Lani, A.}, {Scolini, C.}, {et~al.} 2020, J. Space Weather Space
  Clim., 10, 57

\bibitem[{{Pomoell} \& {Poedts}(2018)}]{Pomoell2018020}
{Pomoell}, J. \& {Poedts}, S. 2018, J. Space Weather Space Clim., 8, A35

\bibitem[{{Powell} {et~al.}(1999){Powell}, {Roe}, {Linde}, {Gombosi}, \& {de
  Zeeuw}}]{Powell1999}
{Powell}, K.~G., {Roe}, P.~L., {Linde}, T.~J., {Gombosi}, T.~I., \& {de Zeeuw},
  D.~L. 1999, J. Comput. Phys., 154, 284

\bibitem[{Priest(2014)}]{priest_2014}
Priest, E. 2014, Magnetohydrodynamics of the Sun (Cambridge: Cambridge
  University Press), 144--176

\bibitem[{{Regnault} {et~al.}(2023){Regnault}, {Strugarek}, {Janvier},
  {Auchere}, {Lugaz}, \& {Al-Haddad}}]{Regnault_2023}
{Regnault}, F., {Strugarek}, A., {Janvier}, M., {et~al.} 2023, A \& A, 670, A14

\bibitem[{{Reiss} {et~al.}(2019){Reiss}, {MacNeice}, {Mays}, {Arge}, {Mostl},
  {Nikolic}, \& {Amerstorfer}}]{Reiss_2019}
{Reiss}, M.~A., {MacNeice}, P.~J., {Mays}, L.~M., {et~al.} 2019, ApJS, 240, 35

\bibitem[{{Riley} {et~al.}(2012){Riley}, {Linker}, {Lionello}, \&
  {Miki{\'c}}}]{RILEY20121}
{Riley}, P., {Linker}, J.~A., {Lionello}, R., \& {Miki{\'c}}, Z. 2012, J.
  Atmos. Sol.-Terr. Phys., 83, 1

\bibitem[{{Riley} {et~al.}(2015){Riley}, {Lionello}, {Linker}, {Cliver},
  {Balogh}, {Beer}, {Charbonneau}, {Crooker}, {DeRosa}, {Lockwood}, {Owens},
  {McCracken}, {Usoskin}, \& {Koutchmy}}]{Riley_2015}
{Riley}, P., {Lionello}, R., {Linker}, J.~A., {et~al.} 2015, ApJ, 802, 105

\bibitem[{{Riley} {et~al.}(2011){Riley}, {Lionello}, {Linker}, {Miki{\'c}},
  {Luhmann}, \& {Wijaya}}]{Riley2011}
{Riley}, P., {Lionello}, R., {Linker}, J.~A., {et~al.} 2011, Sol. Phys., 274,
  361

\bibitem[{Samara {et~al.}(2021)Samara, Pinto, Magdaleni, Wijsen, Jeri, Scolini,
  Jebaraj, Rodriguez, \& Poedts}]{Samara_2021}
Samara, E., Pinto, R.~F., Magdaleni, J., {et~al.} 2021, A \& A, 648, A35

\bibitem[{{Schatten} {et~al.}(1969){Schatten}, {Wilcox}, \&
  {Ness}}]{Schatten1969SoPh}
{Schatten}, K.~H., {Wilcox}, J.~M., \& {Ness}, N.~F. 1969, Sol. Phys., 6, 442

\bibitem[{{Scolini, C.} {et~al.}(2019){Scolini, C.}, {Rodriguez, L.}, {Mierla,
  M.}, {Pomoell, J.}, \& {Poedts, S.}}]{Scolini2019}
{Scolini, C.}, {Rodriguez, L.}, {Mierla, M.}, {Pomoell, J.}, \& {Poedts, S.}
  2019, A \& A, 626, A122

\bibitem[{{Sharov} {et~al.}(2000){Sharov}, {Luo}, {Baum}, \&
  {Loehner}}]{Sharov2000}
{Sharov}, D., {Luo}, H., {Baum}, J., \& {Loehner}, R. 2000, in 38th Aerospace
  Sciences Meeting and Exhibit, aIAA 2000-927

\bibitem[{{Shen} {et~al.}(2021){Shen}, {Liu}, \& {Yang}}]{Shen_2021}
{Shen}, F., {Liu}, Y.~S., \& {Yang}, Y. 2021, ApJS, 253, 12

\bibitem[{{Singer} {et~al.}(2001){Singer}, {Heckman}, \& {Hirman}}]{Singer2001}
{Singer}, H.~J., {Heckman}, G.~R., \& {Hirman}, J.~W. 2001, Space Weather
  Forecasting: {A} Grand Challenge (Washington, DC: American Geophysical Union
  (AGU)), 23--29

\bibitem[{{Singh} {et~al.}(2020){Singh}, {Yalim}, {Pogorelov}, \&
  {Gopalswamy}}]{Singh_2020}
{Singh}, T., {Yalim}, M.~S., {Pogorelov}, N.~V., \& {Gopalswamy}, N. 2020, ApJ,
  894, 49

\bibitem[{{Siscoe}(2000)}]{Siscoe20001223}
{Siscoe}, G. 2000, J. Atmos. Sol.-Terr. Phys., 62, 1223

\bibitem[{{Sitaraman} \& {Raja}(2013)}]{SITARAMAN2013364}
{Sitaraman}, H. \& {Raja}, L.~L. 2013, J. Comput. Phys., 251, 364

\bibitem[{{Steinolfson} \& {Hundhausen}(1988)}]{Steinolfson1988}
{Steinolfson}, R.~S. \& {Hundhausen}, A.~J. 1988, J. Geophys. Res.: Space
  Phys., 93, 14269

\bibitem[{{Suess} {et~al.}(1996){Suess}, {Wang}, \& {Wu}}]{Suess1996}
{Suess}, S., {Wang}, A.-H., \& {Wu}, S. 1996, J. Geophys. Res.: Space Phys.,
  101, 19957

\bibitem[{{Sun} {et~al.}(2011){Sun}, {Liu}, {Hoeksema}, {Hayashi}, \&
  {Zhao}}]{Sun2011}
{Sun}, X., {Liu}, Y., {Hoeksema}, J.~T., {Hayashi}, K., \& {Zhao}, X. 2011,
  Sol. Phys., 270, 9

\bibitem[{{Tanaka}(1995)}]{Tanaka1995}
{Tanaka}, T. 1995, J. Geophys. Res.: Space Phys., 100, 12057

\bibitem[{Thompson \& Reginald(2008)}]{Thompson2008}
Thompson, W. \& Reginald, N. 2008, Sol. Phys., 250, 443

\bibitem[{{Thompson}(2006)}]{Thompson2006CoordinateSF}
{Thompson}, W.~T. 2006, A \& A, 449, 791

\bibitem[{{Thompson} {et~al.}(2003){Thompson}, {Davila}, {Fisher}, {Orwig},
  {Mentzell}, {Hetherington}, {Derro}, {Federline}, {Clark}, {Chen},
  {Tveekrem}, {Martino}, {Novello}, {Wesenberg}, {StCyr}, {Reginald}, {Howard},
  {Mehalick}, {Hersh}, {Newman}, {Thomas}, {Card}, \&
  {Elmore}}]{Thompson2003COR1IC}
{Thompson}, W.~T., {Davila}, J.~M., {Fisher}, R.~R., {et~al.} 2003, in SPIE
  Astronomical Telescopes + Instrumentation

\bibitem[{Titov {et~al.}(2017)Titov, Downs, Miki{\'c}, T{\"o}r{\"o}k, Linker,
  \& Caplan}]{Titov_2018}
Titov, V.~S., Downs, C., Miki{\'c}, Z., {et~al.} 2017, ApJL, 852

\bibitem[{Titov {et~al.}(2014)Titov, T{\"o}r{\"o}k, Miki{\'c}, \&
  Linker}]{Titov_2014}
Titov, V.~S., T{\"o}r{\"o}k, T., Miki{\'c}, Z., \& Linker, J.~A. 2014, ApJ,
  790, 163

\bibitem[{{T{\"o}r{\"o}k, T.} {et~al.}(2010){T{\"o}r{\"o}k, T.}, {Berger, M.
  A.}, \& {Kliem, B.}}]{TOROK2010}
{T{\"o}r{\"o}k, T.}, {Berger, M. A.}, \& {Kliem, B.} 2010, A \& A, 516, A49

\bibitem[{T{\' o}th(2000)}]{Toth2000605}
T{\' o}th, G. 2000, J. Comput. Phys., 161, 605

\bibitem[{{T\'{o}th} {et~al.}(2012){T\'{o}th}, {van der Holst}, {Sokolov}, {De
  Zeeuw}, {Gombosi}, {Fang}, {Manchester}, {Meng}, {Najib}, {Powell}, {Stout},
  {Glocer}, {Ma}, \& {Opher}}]{TOTH2012870}
{T\'{o}th}, G., {van der Holst}, B., {Sokolov}, I.~V., {et~al.} 2012, J.
  Comput. Phys., 231, 870

\bibitem[{{Usmanov}(1993)}]{Usmanov1993}
{Usmanov}, A.~V. 1993, Sol. Phys., 146, 377

\bibitem[{{Usmanov} \& {Goldstein}(2003)}]{Usmanov2003}
{Usmanov}, A.~V. \& {Goldstein}, M.~L. 2003, AIP Conf. Proc., 679, 393

\bibitem[{Vourlidas {et~al.}(2012)Vourlidas, Lynch, Howard, \&
  Li}]{Vourlidas2013}
Vourlidas, A., Lynch, B., Howard, R., \& Li, Y. 2012, Sol. Phys., 284

\bibitem[{Vourlidas {et~al.}(2019)Vourlidas, Patsourakos, \&
  Savani}]{Vourlidas2019}
Vourlidas, A., Patsourakos, S., \& Savani, N.~P. 2019, Philos. Trans. R. Soc.,
  A, 377, 20180096

\bibitem[{{Wang} {et~al.}(2022{\natexlab{a}}){Wang}, {Xiang}, {Liu}, {Lv}, \&
  {Shen}}]{Wang_2022}
{Wang}, H.~P., {Xiang}, C.~Q., {Liu}, X.~J., {Lv}, J.~K., \& {Shen}, F.
  2022{\natexlab{a}}, ApJ, 935, 46

\bibitem[{{Wang} {et~al.}(2022{\natexlab{b}}){Wang}, {Zhao}, {Lv}, \&
  {Liu}}]{Wang2022_CJG}
{Wang}, H.~P., {Zhao}, J.~M., {Lv}, J.~K., \& {Liu}, X.~J. 2022{\natexlab{b}},
  Chin. J. Geophys., 65, 2779

\bibitem[{{Wang} {et~al.}(2019){Wang}, {Feng}, \& {Xiang}}]{WANG201967}
{Wang}, Y., {Feng}, X.~S., \& {Xiang}, C.~Q. 2019, Comput. Fluids, 179, 67

\bibitem[{{Wang} {et~al.}(2007){Wang}, N.~R.~{Sheeley}, \& {Rich}}]{Wang_2007}
{Wang}, Y.~M., N.~R.~{Sheeley}, J., \& {Rich}, N.~B. 2007, ApJ, 658, 1340

\bibitem[{{Wu} \& {Dryer}(2015)}]{WuShiTsan2015}
{Wu}, S.~T. \& {Dryer}, M. 2015, Sci. China Earth Sci., 58, 839

\bibitem[{Wu {et~al.}(1999)Wu, Guo, Michels, \& Burlaga}]{Wu1999}
Wu, S.~T., Guo, W.~P., Michels, D.~J., \& Burlaga, L.~F. 1999, J. Geophys.
  Res.: Space Phys., 104, 14789

\bibitem[{Xia {et~al.}(2018)Xia, Teunissen, Mellah, Chan{\'e}, \&
  Keppens}]{Xia_2018}
Xia, C., Teunissen, J., Mellah, I.~E., Chan{\'e}, E., \& Keppens, R. 2018,
  ApJS, 234

\bibitem[{{Xia} {et~al.}(2014){Xia}, {Luo}, \& {Nourgaliev}}]{XIA2014406}
{Xia}, Y.~D., {Luo}, H., \& {Nourgaliev}, R. 2014, Comput. Fluids, 96, 406

\bibitem[{{Xu} {et~al.}(2020){Xu}, {Zhu}, \& {Guo}}]{Xu_2020}
{Xu}, Y., {Zhu}, J.~H., \& {Guo}, Y. 2020, ApJ, 892, 54

\bibitem[{{Yang} {et~al.}(2011){Yang}, {Feng}, {Xiang}, {Zhang}, \&
  {Wu}}]{Yang2011}
{Yang}, L.~P., {Feng}, X.~S., {Xiang}, C.~Q., {Zhang}, S.~H., \& {Wu}, S.~T.
  2011, Sol. Phys., 271, 97

\bibitem[{{Yang} {et~al.}(2021){Yang}, {Wang}, {Feng}, {Xiong}, {Zhang}, {Zhu},
  {Li}, Zhou, {Shen}, {Zhao}, \& {Liu}}]{Yang_2021}
{Yang}, L.~P., {Wang}, H.~P., {Feng}, X.~S., {et~al.} 2021, ApJ, 918, 31

\bibitem[{{Yang} {et~al.}(2017){Yang}, {Feng}, \& {Jiang}}]{YANG2017561}
{Yang}, Y., {Feng}, X.~S., \& {Jiang}, C.~W. 2017, J. Comput. Phys., 349, 561

\bibitem[{{Yang} {et~al.}(2018){Yang}, {Shen}, {Yang}, \&
  {Feng}}]{Yangzicai2018}
{Yang}, Z.~C., {Shen}, F., {Yang}, Y., \& {Feng}, X.~S. 2018, Chin. J. Space
  Sci., 38, 285

\bibitem[{{Zahr} \& {Persson}(2013)}]{Zahr2013}
{Zahr}, M.~J. \& {Persson}, P.-O. 2013, in 21st AIAA Computational Fluid
  Dynamics Conference, June 24-27, 2013, San Diego, CA.

\bibitem[{Zhang {et~al.}(2012)Zhang, Cheng, \& Ding}]{Zhang2012}
Zhang, J., Cheng, X., \& Ding, M. 2012, Nat. Commun., 3, 747

\bibitem[{Zhang {et~al.}(2023)Zhang, Feng, Li, Xiong, Shen, Yang, Zhao, Zhou,
  \& Liu}]{Zhang2023}
Zhang, M., Feng, X.~S., Li, H.~C., {et~al.} 2023, Front. astron. space sci., 10

\bibitem[{Zhang {et~al.}(2019)Zhang, Feng, \& Yang}]{Zhang2019}
Zhang, M., Feng, X.~S., \& Yang, L.~P. 2019, J. Space Weather Space Clim., 9,
  A33

\bibitem[{{Zhang} {et~al.}(2006){Zhang}, {John Yu}, {Henry Lin}, Chang, \&
  Blankson}]{zhang20060520}
{Zhang}, M.~J., {John Yu}, S.~T., {Henry Lin}, S.~C., Chang, S.-C., \&
  Blankson, I. 2006, J. Comput. Phys., 214, 599

\bibitem[{Zhao {et~al.}(2002)Zhao, Plunkett, \& Liu}]{Zhao2002}
Zhao, X.~P., Plunkett, S.~P., \& Liu, W. 2002, J. Geophys. Res.: Space Phys.,
  107, SSH 13

\bibitem[{{Zhou} \& {Feng}(2014)}]{Zhou2014}
{Zhou}, Y.~F. \& {Feng}, X.~S. 2014, Sci. China Earth Sci., 57, 153

\bibitem[{{Zhou} \& {Feng}(2017)}]{Zhouyufen2017}
{Zhou}, Y.~F. \& {Feng}, X.~S. 2017, J. Geophys. Res.: Space Phys., 122, 1451

\bibitem[{Zhou {et~al.}(2012)Zhou, Feng, Wu, Du, Shen, \& Xiang}]{Zhou2012}
Zhou, Y.~F., Feng, X.~S., Wu, S.~T., {et~al.} 2012, J. Geophys. Res.: Space
  Phys., 117

\end{thebibliography}
\clearpage

\begin{appendix}
\section{Orszag-Tang MHD vortex problem}\label{sec:Orszag–Tang MHD Vortex Problem}
    The Orszag-Tang vortex system includes many significant characteristics of MHD turbulence, and some shocks and other discontinuities occur in this MHD vortex flow as time progresses \citep{orszag_tang_1979}. In this section, we simulate the Orszag-Tang vortex flow by the ERK2 described in Eq. (\ref{2orderRK}) and the pseudo-time marching method (P-t) described in Sect. \ref{sec: Pseudotimemarching}, respectively.

   The computational domain of the Orszag-Tang vortex system is set as $[0, 2\pi]\times [0,2\pi]$, and the periodic boundary condition is implemented in both $x$ and $y$ directions. The initial state is:
$$
\begin{aligned}
&\rho\left(x,y\right)=\gamma^2, \quad u\left(x,y\right)=-\sin y,\\
&v\left(x,y\right)=\sin x, \quad w\left(x,y\right)=0,\\
&p\left(x,y\right)=\gamma, \quad B_x\left(x,y\right)=-\sin y,\\
&B_y\left(x,y\right)=\sin 2x, \quad B_z\left(x,y\right)=0,
\end{aligned}
$$
with $\gamma=\frac{5}{3}$.
   The grid resolution of the Orszag-Tang vortex flow modelled by the ERK2 scheme is $400 \times 400$, and $\rm CFL$ is set to be 0.5. As for the simulation using the P-t method, we adopt grid resolutions of $400 \times 400$ and $800 \times 800$ and set $\rm CFL$ to 2 and 4, respectively. During the simulation, the physical time steps are around $3\times 10^{-3}$ for the test modelled by the ERK2 scheme and $1.2\times 10^{-2}$ for the tests modelled by the P-t method. Since $\Delta t$ is not too large and the matrix $\left(\frac{\partial{\mathbf{R}^{'}}}{\partial \mathbf{U}}\right)^{n+1,m}$ in Eq. (\ref{modifiedglobalPseIBElinearized}) does not change obviously in the P-t iterations of a physical time step, we modify Eq. (\ref{modifiedglobalPseIBElinearized}) as bellow, inspired by the Jacobian recycling strategies proposed in \cite{PERSSON2013414} and \cite{Zahr2013}, to reduce the computation cost.
\begin{equation}\label{linearizedBackEulerpsedotimeMF}
\left(\frac{V}{\Delta \tau}\mathbf{I}+\frac{V}{\Delta t}\mathbf{I}+\left(\frac{\partial{\mathbf{R}^{'}}}{\partial \mathbf{U}}\right)^{n+1,0}\right)\Delta \mathbf{U}^{n+1,m}=\frac{V}{\Delta t}\left(\mathbf{U}^n-\mathbf{U}^{n+1,m}\right)-\mathbf{R}^{n+1,m}
,\end{equation}

\begin{figure*}[htpb!]
\begin{center}
  \vspace*{0.01\textwidth}
    \includegraphics[width=0.8\linewidth,trim=1 1 1 1, clip]{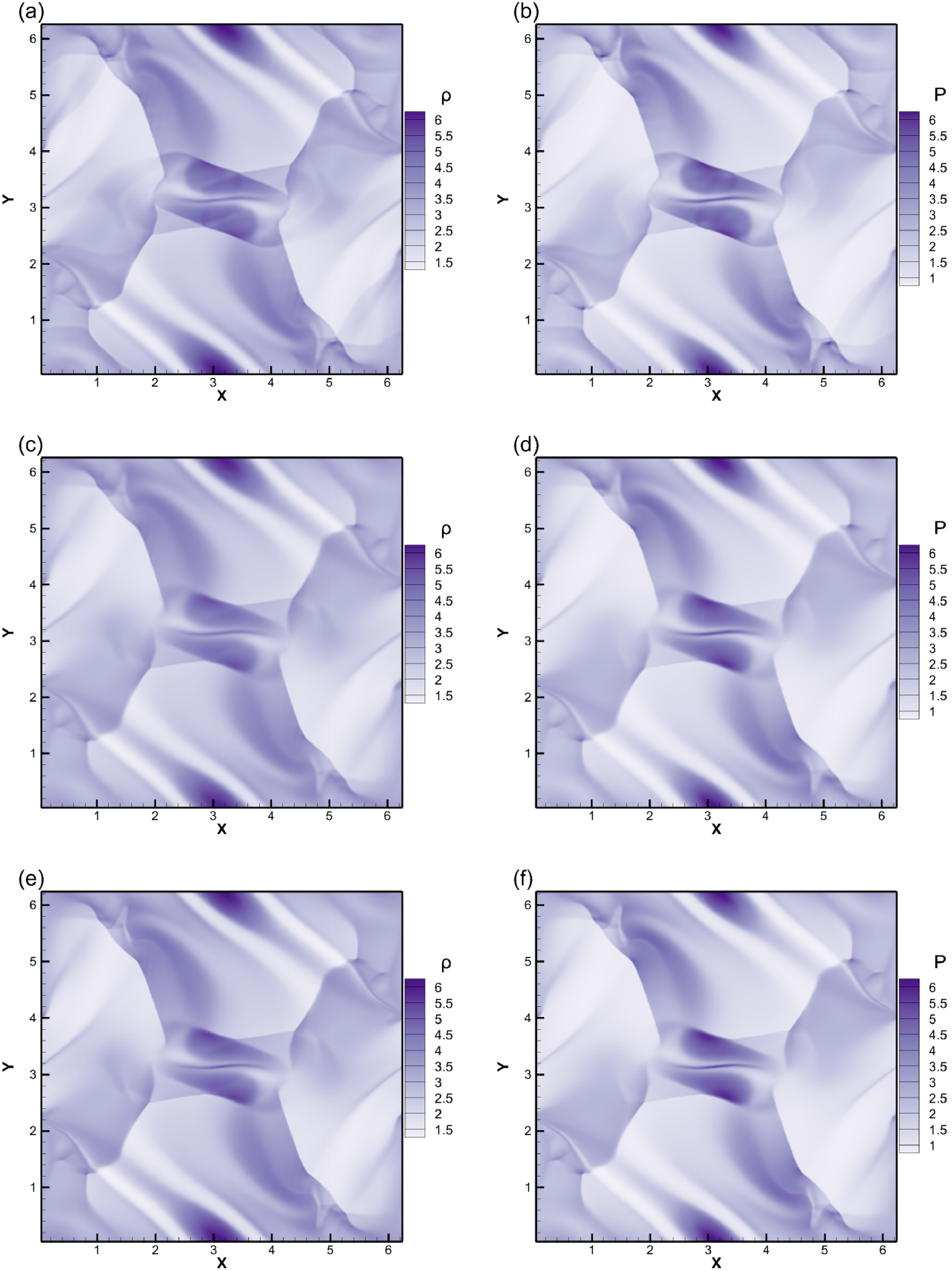}
\end{center}
\caption{Contours of density (a, c, and e) and thermal pressure (b, d, and f) for the MHD vortex problem at $t = 3$, modelled by ERK2 with 400 $\times$ 400 grids (a, b),  by P-t with 400 $\times$ 400 grids (c, d), and by P-t with 800 $\times$ 800 grids (e, f).}\label{VortexrhoandpCONTOUR}
\end{figure*}
\begin{figure*}[htpb!]
\begin{center}
  \vspace*{0.01\linewidth}
    \includegraphics[width=0.8\textwidth,trim=1 1 1 1, clip]{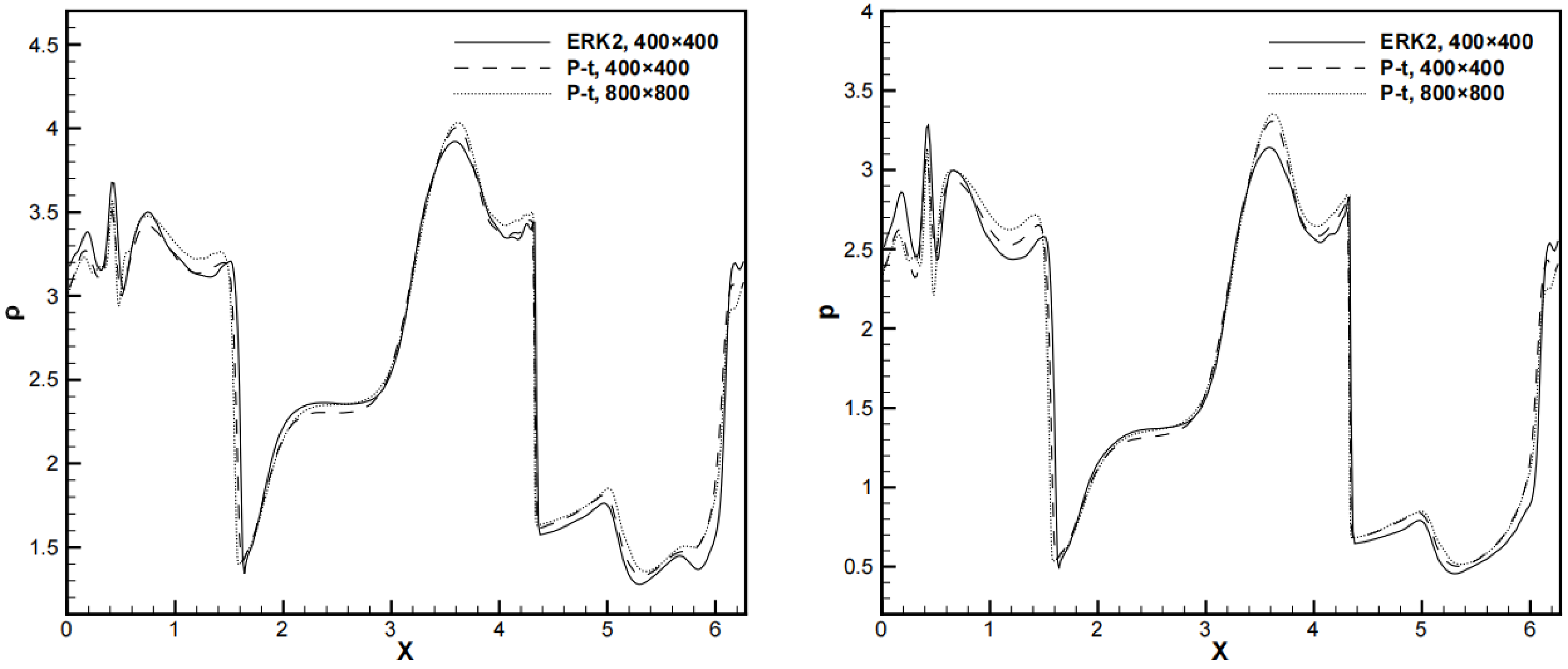}
\end{center}
\caption{Profiles of density (left) and thermal pressure (right) at time $t = 3$ for the MHD vortex problem along line $y = 0.625\pi$, modelled by ERK2 with 400 $\times$ 400 grids (solid lines),  by P-t with 400 $\times$ 400 grids (dashed lines), and by P-t with 800 $\times$ 800 grids (dotted lines).}\label{Vortexrhoandpalongx}
\end{figure*}
  
   In Fig.~\ref{VortexrhoandpCONTOUR}, the contour images of density and thermal pressure of the Orszag-Tang vortex flow at $t = 3$, modelled by the ERK2 scheme with a resolution of  400 $\times$ 400 grid cells, and modelled by the P-t method with resolutions of 400 $\times$ 400 grid cells and 800 $\times$ 800 grid cells, are illustrated. The CFL number of the physical time step is set to be 0.5, 2, and 4, respectively. All these flow fields evolve in symmetrical patterns, and more detailed structures of low density are identified by the P-t method with the refined mesh. In Fig.~\ref{Vortexrhoandpalongx}, we compare the density and thermal pressure profiles along the $y = 0.625\pi$ line at $t = 3$. It shows that shock discontinuities are formed around $x=0.5$, $x=1.6$, and $x=4.4$ for all these three tests, and the shock discontinuities modelled by P-t method with the refined mesh are sharper than those modelled by the ERK2 scheme with coarse mesh.

   It can be seen that these modelled results conform to the previous simulations \citep{Balsara20101970,FengandLiu2019,Fuchs2009,Jiang2010,YANG2017561}, demonstrating that the implicit LU-SGS method and pseudo-time marching method used in this solar coronal model can also be used to simulate small-scale unsteady flows accurately. Therefore, the MHD coronal model proposed in this paper can indeed capture such small-scale MHD flow structures.

\section{Data communication between different components}\label{sec:data communication between different components}
\begin{figure*}[htpb]
\begin{center}
  \vspace*{0.01\textwidth}
    \includegraphics[width=0.7\linewidth,trim=1 1 1 1, clip]{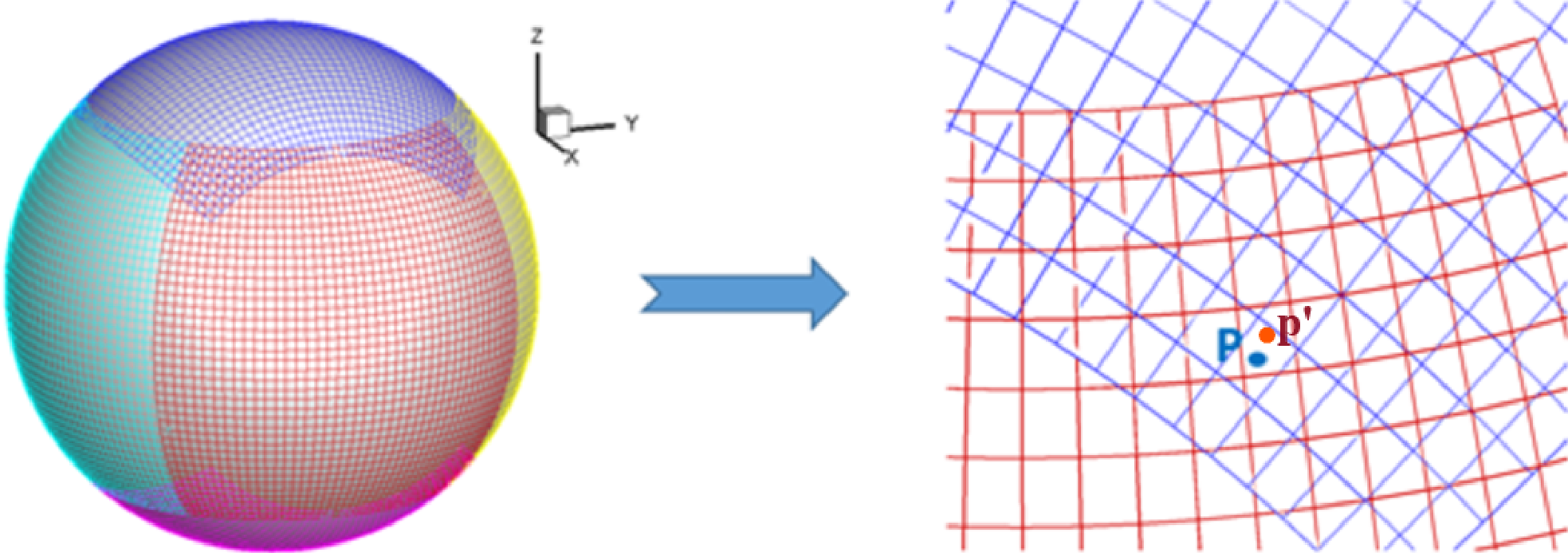}
\end{center}
\caption{Illustration of the data communication between different components. The point denoted by P is a centroid of the ghost cell of the component with blue
  grids and is also in the computation domain of an adjacent component with red grids. The point denoted by $\rm P^{'}$ is the centroid of the cell closest to P in the component with red grids. The reconstruction formulation of a variable is first calculated in the stencil centred on a cell with its centroid denoted by $\rm P^{'}$ in the component with red grids and then transferred to the ghost cell with its centroid denoted by P in the component with blue grids.}\label{Grid_Interpolation}
\end{figure*}

   To improve the precision of data communication, we transfer the reconstructed formulation of variables, not just the point values, between adjacent processors whose grid meshes share some overlapping area.
   As illustrated in Fig.~\ref{Grid_Interpolation}, we first derive the reconstruction formulation of a variable in the ghost cell of the blue component, the centroid of this ghost cell is denoted by P, from the stencil in red component, and then send this reconstruction formulation to the blue component to provide solution information in the ghost cell of this blue component.

   During this data communication procedure, we first search for the cell centroid $\rm P^{'}$, which is the closest to P in the component with red grids. The cell $P'$ with centroid denoted by $\rm P^{'}$ and its six neighbouring cells which share an interface with cell $P'$ serve as a stencil. We implement the RBF interpolation method \citep{LIU20161096,Wang_2022} to calculate the variable at point P. Afterwards, we calculate a second-order Taylor polynomial expanding from P in the component with red grids by employing a least-square (LSQ) method \citep{BARTH1991,BARTH1993}, while the stencil consists of point P, cell $P'$ and the six neighbouring cells that share an interface with cell $P'$. Finally, the second-order Taylor polynomial derived from the red component is sent to the blue component to maintain synchronisation of this blue component's ghost and inner cells. This synchronised MPI data communication is implemented before each LU-SGS iteration in both the quasi-steady coronal simulations and time-dependent CME simulations to help maintain the synchronisation of each processor's ghost cells and inner cells.
\end{appendix}

\end{document}